\documentclass{ametsocV6.1}

\usepackage{booktabs}
\usepackage{tabularx}
\usepackage{amssymb}
\usepackage{pifont}
\usepackage{xcolor}
\usepackage{rotating}
\usepackage{float}
\usepackage{graphicx}
\usepackage{adjustbox}
\usepackage{nameref}

\makeatletter
\let\internallinenumbers\relax
\makeatother

\newcommand{\cmark}{\raisebox{0.1ex}{\color[HTML]{006400}\ding{51}}} 
\newcommand{\xmark}{\raisebox{0.1ex}{\color[HTML]{FF6347}\ding{55}}} 

\title{MSWEP~V3: Machine Learning-Powered Global Precipitation Estimates at 0.1° Hourly Resolution (1979–Present)}

\authors{
Xuetong Wang,\aff{a}
Raied S. Alharbi,\aff{b} 
Oscar M. Baez-Villanueva,\aff{c} 
Diego G. Miralles,\aff{c} 
Jun Ma,\aff{a}
Shiqin Xu,\aff{a}
Matthew F. McCabe,\aff{a}
Florian Pappenberger,\aff{d}
Albert I.J.M. van Dijk,\aff{e} 
Tim R. McVicar,\aff{f}
Lanka Karthikeyan,\aff{g,h}
Hayley J. Fowler,\aff{i}
Ming Pan,\aff{j}
Solomon H. Gebrechorkos,\aff{k,l} Hylke E.~Beck,\aff{a}\correspondingauthor{Hylke E.~Beck, hylke.beck@kaust.edu.sa}}

\affiliation{
\aff{a}{King Abdullah University of Science and Technology (KAUST), Thuwal, Saudi Arabia}\\
\aff{b}{Department of Civil Engineering, College of Engineering, King Saud University, Riyadh, Saudi Arabia}\\
\aff{c}{Hydro-Climate Extremes Lab (H-CEL), Ghent University, Ghent, Belgium}\\
\aff{d}{ECMWF, Reading, United Kingdom}\\
\aff{e}{Fenner School of Environment \& Society, Australian National University, Canberra, ACT, Australia}\\
\aff{f}{CSIRO Environment, Canberra, Australian Capital Territory, Australia}\\
\aff{g}{Centre of Studies in Resources Engineering, IIT Bombay, Mumbai, India}\\
\aff{h}{Centre for Climate Studies, IIT Bombay,Mumbai, India}\\
\aff{i}{School of Engineering, Newcastle University, Newcastle upon Tyne, UK}\\
\aff{j}{Center for Western Weather and Water Extremes, Scripps Institution of Oceanography, University of California, San Diego, San Diego, California}\\
\aff{k}{School of Geography and the Environment, University of Oxford, Oxford, UK}\\
\aff{l}{School of Geography and Environmental Science, University of Southampton, Southampton, UK}
}

    \abstract{We introduce Version~3 (V3) of the gridded near real-time Multi-Source Weighted-Ensemble Precipitation (MSWEP) product---the first fully global, historical machine learning-powered precipitation ($P$) dataset, developed to meet the growing demand for timely and accurate $P$ estimates amid escalating climate challenges. MSWEP~V3 provides hourly data at 0.1$^\circ$ resolution from 1979 to the present, continuously updated with a latency of approximately two hours. Development follows a two-stage process. First, baseline $P$ fields are generated using machine learning model stacks that integrate satellite- and (re)analysis-based $P$ and air-temperature products, along with static variables. The models are trained using hourly and daily observations from 15,959 $P$ gauges worldwide. Second, these baseline $P$ fields are corrected using daily and monthly gauge observations from 57,666 and 86,000 stations globally. To assess MSWEP~V3's baseline performance, we evaluated 19 (quasi-) global gridded $P$ products---including both uncorrected and gauge-based products---using observations from an independent set of 15,958 gauges excluded from the first training stage. The MSWEP~V3 baseline achieved a median daily Kling-Gupta Efficiency (KGE) of 0.69, outperforming all evaluated products. Other uncorrected products achieved median daily KGE values of 0.61 (ERA5), 0.46 (IMERG-L V7), 0.38 (GSMaP V8), and 0.31 (CHIRP). Using leave-one-out cross-validation, the daily gauge correction was found to improve the median daily correlation by 0.09, constrained by the already strong baseline performance. We anticipate that MSWEP~V3---accessible at \url{www.gloh2o.org/mswep}---will enable more reliable monitoring, forecasting, and management of water-related risks in a variable and changing climate.}

\begin{document}

\maketitle
\nolinenumbers


\section{Introduction}   

The global hydrological cycle is intensifying, resulting in increases in both precipitation ($P$) extremes and drought severity at global and regional scales \citep{huntington_evidence_2006,zhang_future_2019,allan_advances_2020,de_luca_concurrent_2020,gebrechorkos_warming_2025}. Furthermore, anthropogenic warming increases atmospheric moisture capacity which in turn alters storm dynamics, leading to more frequent and intense $P$ events \citep{fischer_anthropogenic_2015}. At the same time, many regions face growing water scarcity driven by rising atmospheric water demand, accelerated evaporation, and shifting $P$ patterns \citep{hanjra_global_2010,ingrao_water_2023}. Together, these pressures exacerbate the risks associated with floods and droughts, highlighting the need for accurate, timely, and high-resolution $P$ data to support a wide range of critical applications, including hydrological modeling, flood and drought monitoring and forecasting, proactive agricultural practices, water resource management, and climate change attribution \citep{brunner_challenges_2021,jahanddideh-tehrani_role_2021}. To inform such efforts, numerous gridded $P$ products have been developed over the last few decades based on satellite retrievals (e.g., TRMM, GPM), (re)analysis outputs (e.g., ERA5, JRA-3Q), gauge observations, or combinations thereof (see, e.g., \citealp{sun_review_2018,serrano-notivoli_rain_2021,abbas_comprehensive_2025}). Often, these products are designed with specific priorities in mind, balancing record length, temporal homogeneity, instantaneous accuracy, spatial or temporal resolution, or latency.

The Multi-Source Weighted-Ensemble Precipitation (MSWEP) product was first released in 2017 and quickly became one of the most widely-used gridded $P$ products. It has supported a broad spectrum of hydrological and climatological applications, including flood modeling (e.g., \citealp{grenier_flood_2024}), drought monitoring (e.g., \citealp{gebrechorkos_global_2023}), land surface modeling (e.g., \citealp{shrestha_role_2020}), and water resource assessments (e.g., \citealp{lakew_evaluation_2020}). MSWEP has shown strong performance relative to other gridded $P$ products in several global evaluations (e.g., \citealp{beck_global-scale_2017, gebrechorkos_global-scale_2024, abbas_comprehensive_2025}) and numerous regional evaluations (e.g., \citealp{beck_daily_2019, sharifi_performance_2019, ali_long-term_2022, wang_saudi_2025}). Its key features include: (i)~integration of satellite retrievals and (re)analysis outputs to exploit their complementary strengths in capturing convective and frontal $P$, respectively; (ii)~daily gauge corrections that explicitly account for reporting times, enhancing performance in regions with dense observational networks; (iii)~high 0.1$^\circ$ spatial and hourly temporal resolution (3-hourly for Version~2, V2), with full global coverage; and (iv)~a long-term record extending from 1979 to the present with a latency of $\sim$2~hours (3--4~hours for~V2), enabling both retrospective analysis and operational applications.

Despite these advantages, the previous MSWEP version (V2; \citealp{beck_mswep_2019})---superseded by V3 presented here---still showed some limitations that affect its performance in specific regions and applications. First, it tends to underestimate peak $P$ intensities, reducing its ability to capture extreme events (e.g., \citealp{satge_consistency_2019,dong_evaluation_2022,li_evaluation_extreme_2022}). Second, it exhibits an overabundance of low-intensity drizzle events, affecting $P$ frequency statistics and water balance estimates (e.g., \citealp{sahlu_evaluation_2017,bai_which_2020,yang_reliability_2020}). Third, the historical record ends in 2021, precluding robust analysis of more recent $P$ events or trends. Fourth, its 3-hour resolution is too coarse to represent the rapid variability of convective $P$ events and limits its usefulness for flash flood and landslide prediction \citep{gonzalez_systematic_2024}. Fifth, and perhaps most importantly, it relies on a relatively simple statistical merging scheme with static weight maps that do not adequately capture spatially and temporally varying differences in input performance.

Machine learning (ML) has demonstrated its potential as a powerful approach for improving $P$ estimation. A wide range of ML models, typically trained on gauge observations, have been proposed---from more classical multivariate linear regression (MLR), artificial neural networks (ANNs), support vector machines (SVMs), and decision trees, to modern deep learning architectures such as Convolutional Neural Networks (CNNs), Long Short-Term Memory (LSTM) networks, transformers, Graph Neural Networks (GNNs), and various hybrid approaches (see reviews by \citealp{hussein_rainfall_2022,papacharalampous_comparison_2023,dotse_review_2024,xu_multi-source_2024}). However, most applications to precipitation so far have been confined to small regions or individual catchments, with unknown generalizability of findings. Furthermore, the large majority of studies are methodological and not oriented towards producing publicly accessible gridded $P$ datasets. A notable recent exception is \citet{wang_saudi_2025}, who developed a high-resolution gridded $P$ product for the Arabian Peninsula by stacking XGBoost (XGB; \citealp{chen_xgboost_2016}) and Random Forest (RF; \citealp{breiman_random_2001}) models. Their framework integrated dynamic predictors from satellite and (re)analysis sources---including $P$ and 2-m air temperature ($T$)---alongside static geographic, topographic, and climatic predictors. The resulting product significantly outperformed other gridded $P$ products in the region and was made available for operational and research applications.

This study introduces the next generation of MSWEP (V3) which scales up the ML-based methodology of \citet{wang_saudi_2025} to the global domain while retaining the multi-source basis of previous MSWEP versions to provide $P$ fields with improved accuracy. ML-powered baseline $P$ fields are corrected using daily and monthly gauge observations through a method that explicitly accounts for gauge proximity and reporting times and, new in V3, considers spatial $P$ correlation lengths and inter-gauge dependencies. MSWEP~V3 features an hourly temporal resolution (an improvement over the 3-hourly resolution of~V2) from 1979 to approximately 2~hours before the present (3--4~hours for~V2), at a spatial resolution of 0.1$^\circ$ (about 11~km at the equator). 
 
\section{Data and Methods}
\label{DataMethods}

\subsection{Predictors} 

MSWEP~V3 baseline $P$ fields are generated from 1979 onward using a combination of dynamic and static predictors. The dynamic (time-varying) predictors include satellite products (IMERG-Early/Late~V07, GSMaP-Std/NRT~V8, PERSIANN-CCS-CDR and PDIR-Now) and (re)analysis products (ERA5 and GDAS; Table~\ref{globalPProductDetails}). IMERG-Early/Late~V07 \citep{huffman_gpm_2019} and GSMaP-Std/NRT~V8 \citep{kubota_global_2020} provide $P$ data from 2000 onward, available up to $\sim$3--4 hours before real time. Prior to 2000, we use PERSIANN-CCS-CDR \citep{sadeghi_persiann-ccs-cdr_2021}, while PDIR-Now \citep{nguyen_persiann_2020} is used beyond the IMERG/GSMaP latency window. ERA5 \citep{hersbach_era5_2020} and GDAS \citep{noaa_ncep_national_2025} provide additional $P$ inputs and daily mean $T$ to capture seasonal error patterns.


We also include several static predictors selected for their relevance to precipitation modeling: two climate indices—Aridity Index (AI) and mean annual $P$ (Pmean); one topographic metric—Effective Terrain Height (ETH); and three geographic variables—latitude (Lat), longitude (Lon), and absolute latitude (AbsLat; Appendix Table~\ref{StatPredictor}). For more details on the dynamic and static predictors, see Appendix~A ``\nameref{DataPredictors}''.

\begin{sidewaystable}[h!]
\footnotesize
\caption{Overview of quasi- and fully-global products used in this study. Spatial coverage is denoted as ``Global'' for full global coverage (including oceans) and ``Land'' for land-only products. ``Utilization'' indicates whether a product was used only for evaluation or also as a dynamic predictor in the ML models for generating MSWEP~V3 baseline $P$ fields. Version information is unavailable for most products. Abbreviations: $P$ = precipitation; $T$ = temperature; S = satellite; G = gauge; Re = reanalysis; A = analysis; NRT = near-real-time; Pred = used as predictor; Eval = used for evaluation.}

\centering
    \begin{tabular}{llllllllllll}
    \toprule
    &&&&\multicolumn{2}{c}{Resolution}&&\multicolumn{2}{c}{Coverage}\\
    \cline{5-6} \cline{8-9}
    Data & Version & Variables & Data Source &  Temporal & Spatial && Temporal & Spatial &Time Latency& References/URL & Utilization \\
    \midrule
    IMERG-E  & V07 & $P$ & S  & 30 min & 0.1$^\circ$ && 2000--NRT & 60$^\circ$ N/S & $\sim$4 hours& \citet{huffman_gpm_2019} & Eval, Pred \\
    IMERG-L  & V07 & $P$ & S  & 30 min & 0.1$^\circ$ && 2000--NRT & 60$^\circ$ N/S & $\sim$12 hours& \citet{huffman_gpm_2019} & Eval, Pred \\
    IMERG-F  & V07 & $P$ & S, G  & 30 min & 0.1$^\circ$ && 2000--NRT & 60$^\circ$ N/S &$\sim$3 months & \citet{huffman_gpm_2019} & Eval \\
    GSMaP-NRT & V8  & $P$ & S & Hourly & 0.1$^\circ$ && 2000--NRT & 60$^\circ$  N/S & $\sim$4 hours& \citet{kubota_global_2020} & Eval, Pred \\
    GSMaP-MVK & V8  & $P$ & S & Hourly & 0.1$^\circ$ && 2000--NRT & 60$^\circ$  N/S & $\sim$3~days& \citet{kubota_global_2020} & Eval, Pred \\
    ERA5 & / & $P$ & Re  & Hourly & 0.25$^\circ$ && 1940--NRT & Global &$\sim$5 days& \citet{hersbach_era5_2020} & Eval, Pred\\
    ERA5 &/ & $T$ & Re  & Hourly & 0.25$^\circ$ && 1940--NRT & Global &$\sim$5 days& \citet{hersbach_era5_2020} & Pred\\
    GDAS &/ & $P$ & A & Hourly & 0.25$^\circ$ && 2001$^1$--NRT & Global & $\sim$3--6~hours & $^2$ & Eval, Pred \\
    GDAS &/ & $T$ & A & Hourly & 0.25$^\circ$ && 2001$^1$--NRT & Global & $\sim$3--6~hours &$^2$& Pred \\
    PDIR-Now &/ & $P$ & S &  Hourly & 0.04$^\circ$ && 2000--NRT & 60$^\circ$  N/S & $\sim$100~minutes & \citet{nguyen_persiann_2020} & Eval, Pred\\
    PERSIANN-CCS-CDR &/ & $P$ & S, G & 3-hourly & 0.04$^\circ$ && 1983--2021 & 60$^\circ$ N/S &/& \citet{sadeghi_persiann-ccs-cdr_2021} & Eval, Pred\\ 
    JRA-3Q &/ & $P$ & Re & 3-hourly & $\sim$40~km && 1947--NRT & Global &$\sim$3 days & \citet{kosaka_jra-3q_2024} & Eval\\
    CMORPH-RAW &/ & $P$ & S & 30 min & $\sim$8~km && 2019--NRT & 60$^\circ$N/S & $\sim$4 hours& \citet{joyce_cmorph_2004} & Eval\\
    CMORPH-RT &/ & $P$ & S & 30 min & $\sim$8~km && 2019--NRT & 60$^\circ$N/S & $\sim$4 hours& \citet{xie_reprocessed_2017} & Eval\\
    PERSIANN-CCS &/ & $P$ & S & Hourly & 0.04$^\circ$ && 2003--NRT & 60$^\circ$ N/S & $\sim$90~minutes & \citet{hong_precipitation_2004} & Eval\\ 
    CPC Unified &/ & $P$ & G & Daily & 0.5$^\circ$ && 1979--NRT & Land & $\sim$1 day & \citet{chen_assessing_2008} & Eval\\
    SM2RAIN-CCI &/ & $P$ & S & Daily & 0.25$^\circ$ && 1998--2015 & Land & / & \citet{ciabatta_sm2rain-cci_2018} & Eval\\
    SM2RAIN-ASCAT &/ & $P$ & S & Daily & 0.1$^\circ$ && 2007--2021 & Land & / & \citet{brocca_sm2rainascat_2019} & Eval \\
    SM2RAIN-GPM &/& $P$ & S & Daily & 0.25$^\circ$ && 2007--2018 & Land & / & \citet{massari_daily_2020} & Eval \\ 
    CHIRP & V2 & $P$ & S, Re, A & Daily & 0.05$^\circ$ && 1981--NRT& Land, 50$^\circ$ N/S & $\sim$6 days & \citet{funk_climate_2015} & Eval\\
    CHIRPS & V2 & $P$ & S, G, Re & Daily & 0.05$^\circ$ &&1981--NRT& Land, 50$^\circ$ N/S & 2~weeks & \citet{funk_climate_2015} & Eval\\
    MSWEP & V2.8 & $P$ & S, G, Re, A & 3-hourly & 0.1$^\circ$ && 1979--NRT & Global & $\sim$3 hours & \citet{beck_mswep_2019} & Eval\\
    MSWEP & V3.15 & $P$ & S, G, Re, A & Hourly & 0.1$^\circ$ && 1979--NRT & Global & $\sim$2 hours & This study & Eval\\
    \bottomrule
    \end{tabular}

\par\vspace{0.4em}

\raggedright
\footnotesize
$^1$ We only considered data from 2021 onward, when the Global Forecast System (GFS) used to produce GDAS was upgraded to V16.\\
$^2$ \url{www.ncei.noaa.gov/products/weather-climate-models/global-data-assimilation}.

\label{globalPProductDetails}
\end{sidewaystable}

\subsection{Gauge Observations} 

Daily and hourly $P$ gauge observations were compiled to train and validate the ML models and to correct the baseline MSWEP~V3 $P$ fields. For ML training and validation, we used both direct gauge data (GHCN-D, \citealp{menne_overview_2012}; GSOD, \citealp{noaa_ncep_global_2020}; GSDR, \citealp{lewis_gsdr_2019,lewis_quality_2021}; ISD, \citealp{noaa_ncei_global_2021}) and gridded gauge-radar products (EURADCLIM, \citealp{overeem_euradclim_2023}; Stage-IV, \citealp{lin_12_2005}). Gridded datasets were preferred for model training to reduce the point-to-grid scale mismatch \citep{ensor_statistical_2008} and to avoid daily gauge reporting time misalignment \citep{yang_development_2020}. The gridded data were upscaled to 0.1$^\circ$ resolution through spatial averaging. For the daily correction of the baseline fields, only direct gauge observations were used, as the gridded gauge-radar products have limited temporal coverage. For more details on the gauge observations, see Appendix~A ``\nameref{DataPrecipGaugeObs}''.

Time-series of each daily $P$ were screened to remove suspicious gauges. Gauges were excluded if they (i)~lacked any zero-$P$ days, (ii)~reported daily maxima exceeding 1,825~mm~d$^{-1}$, (iii)~had long-term means outside the range 5--10,000~mm~yr$^{-1}$, (iv)~contained too few wet or dry days, or (v)~had less than one year of valid data for ML model training/validation during 2010--2024 or fewer than five years for the daily baseline gauge correction spanning 1979--2024. After applying all filters, 31,917 gauges were retained for ML model training and validation, with half (15,959) used for training and half (15,958) for validation. For the daily gauge correction, 57,666 gauges passed all filters.  Hourly gauge observations (from GSDR, Stage-IV, and EURADCLIM) were not subjected to additional quality control. For more details on the quality control, see Appendix~A ``\nameref{DailyGaugeDataQC}''.


The baseline $P$ fields were corrected at the monthly time scale using the gridded GPCC Full Data Monthly Product V2022 \citep{schneider_evaluating_2017}, which is based on $\sim$86,000 gauges from 1891--2020 at 0.25$^\circ$ resolution. To extend coverage beyond 2020, we incorporated the GPCC First Guess Monthly Product (2013--present, 0.5$^\circ$). To reduce uncertainty, we retained only grid cells containing actual gauges, effectively reconstructing monthly observations while excluding interpolated cells. No additional quality control was applied beyond that performed by the GPCC developers. For more details on the GPCC data, see Appendix~A ``\nameref{MonthlyPrecipData}''.


\subsection{Estimation of Gauge Reporting Times}

Gauge reporting times vary widely across regions and are typically recorded in local morning hours, often several hours offset from UTC \citep{yang_development_2020}. Ignoring these offsets leads to temporal misalignment between gauge and gridded data, impairing both ML model training and gauge correction \citep{beck_mswep_2019,xiang_evaluation_2021}. Reporting times for daily gauges were inferred by shifting baseline hourly $P$ data in 1-hour steps over a ±36-hour window, recalculating daily totals for each shift, and identifying the shift maximizing Spearman correlation with gauge data. The reverse of this shift was taken as the actual reporting time. For more details on the gauge reporting-time estimation, see Appendix~A ``\nameref{MethodsReportingTimes}''.


\subsection{Estimation of Precipitation Correlation Lengths}

The $P$ correlation length describes the spatial scale over which $P$ exhibits statistical dependence and varies with precipitation type and temporal aggregation \citep{mandapaka_analysis_2013}. Robust estimates are required when applying gauge-based corrections to gridded $P$ products \citep{schiemann_geostatistical_2011,schuurmans_automatic_2007,tapia-silva_geostatistics_2024,funk_climate_2015}. For the MSWEP~V3 gauge correction, fixed daily correlation lengths were calculated at 0.1$^\circ$ resolution from the baseline $P$ fields by computing Spearman correlations and fitting a Gaussian decay model using 20 randomly selected neighboring cells within 500~km of each target cell. For more details on the estimation of $P$ correlation lengths, see  Appendix~A ``\nameref{MethodsCorrelationLengths}''.



\subsection{MSWEP~V3 Algorithm}

Here we provide a brief summary of the MSWEP~V3 algorithm; for a full description, see Appendix~A. The MSWEP~V3 algorithm comprises two stages:

\begin{enumerate}

\item \textbf{Baseline Precipitation Estimation Algorithm:} 

MSWEP~V3 extends globally the algorithm of \citet{wang_saudi_2025}, which uses stacked decision tree–based models to estimate baseline $P$ fields from dynamic predictors (e.g., ERA5, IMERG, GSMaP, and PERSIANN-CCS-CDR) and static predictors (related to location, climate, and topography). The model stack architecture is illustrated in Fig.~\ref{Flowchart}, while their spatiotemporal deployment, including the harmonization procedure (see below), is illustrated in Fig.~\ref{Periods}. Each stack includes four submodels: two XGB models \citep{chen_xgboost_2016} for daily and 3-hourly $P$ estimation, an RF model \citep{breiman_random_2001} for correcting variance underestimation using quantile-matched, square-root-transformed 3-hourly data, and a final XGB model for hourly disaggregation. A total of 18 model stacks were trained to reflect differences in dynamic predictor availability, with model\_01 (or model\_05 at high latitudes) designated as reference (Table~\ref{model_predictors_table}). Outputs from non-reference stacks are harmonized to the reference stack via (i)~detrending, (ii)~Cumulative Distribution Function (CDF) matching, and (iii)~trend restoration to prevent temporal discontinuities. All models were trained and validated using 2010–2024 gauge and predictor data. For training, daily $P$ and $T$ predictor means were recalculated to match the reporting-time offset of each daily gauge, whereas hourly $P$ and $T$ predictor data were shifted to match the reporting-time offset of each hourly gauge. For more details on the baseline $P$ estimation algorithm, see Appendix~A ``\nameref{MethodsMLalgorithm}''.

MSWEP-NRT extends the MSWEP~V3 baseline in near real time with~2-hour latency, using data from GDAS, IMERG, GSMaP, and PDIR-Now. Unlike the infrequently executed historical pipeline, MSWEP-NRT runs continuously, updating hourly as new inputs become available. Initial estimates rely on model\_18, which uses PDIR-Now; higher-quality models (e.g., models\_06--17) replace these as GDAS, IMERG, and GSMaP data become available after 3--4~hours. Full consistency with the historical record is achieved once ERA5 data are released after $\sim$5~days, allowing the reference model\_01 (or model\_05 at high latitudes) to be applied. Each file includes a model stack indicator (1–18) to ensure traceability. For more details on the NRT extension, see Appendix~A ``\nameref{MethodsNRTextension}''.

\item \textbf{Gauge Correction Procedure:} 

MSWEP~V3 corrects daily variability in the historical hourly baseline $P$ fields using daily gauge data. While several other datasets also apply daily gauge corrections (e.g., PERSIANN-CCSA, \citealp{boushaki_bias_2009}; GSMaP\_Gauge, \citealp{mega_gauge_2014}; and MERRA-2, \citealp{gelaro_modern-era_2017,reichle_land_2017}), we use a more sophisticated approach based on optimal interpolation (OI; \citealp{gandin_objective_1963,daley_atmospheric_1993,chen_global_2002}). This method is computationally efficient and statistically robust, accounting for gauge proximity, $P$ correlation length, inter-gauge dependence, and reporting-time alignment. For each grid cell, up to four surrounding gauges within a 500-km radius are selected. Baseline and gauge series are rescaled for consistency, and daily corrections are computed after aligning time series using reporting-time offsets. Multiplicative corrections are applied to the hourly data and combined using OI weights. The final corrected time series are computed as weighted combinations of corrected and uncorrected estimates. For more details on the daily gauge correction, see Appendix~A ``\nameref{MethodsDailyGaugeCorrection}''.


In addition to daily correction, MSWEP~V3 applies a monthly correction using gridded GPCC data \citep{schneider_evaluating_2017} rather than direct monthly gauge observations. In contrast to other products (e.g., IMERG-Final; \citealp{huffman_gpm_2019}, WFDE5; \citealp{cucchi_wfde5_2020}, GPCP; \citealp{huffman_new_2023}, and PERSIANN-CCS-CDR; \citealp{sadeghi_persiann-ccs-cdr_2021}), we exclude interpolated GPCC grid cells and retain only cells that contain gauge observations. An OI scheme, similar to the daily correction, is then applied, accounting for $P$ correlation length, gauge proximity, and inter-gauge dependence. A separate long-term bias adjustment corrects mean offsets using monthly data from the four nearest gauges, with weights derived from the OI scheme. For more details on the monthly gauge correction, see Appendix~A ``\nameref{MethodsMonthlyGaugeCorrection}''.

\end{enumerate}

\begin{figure}
\centering
\includegraphics[width=0.75\linewidth]{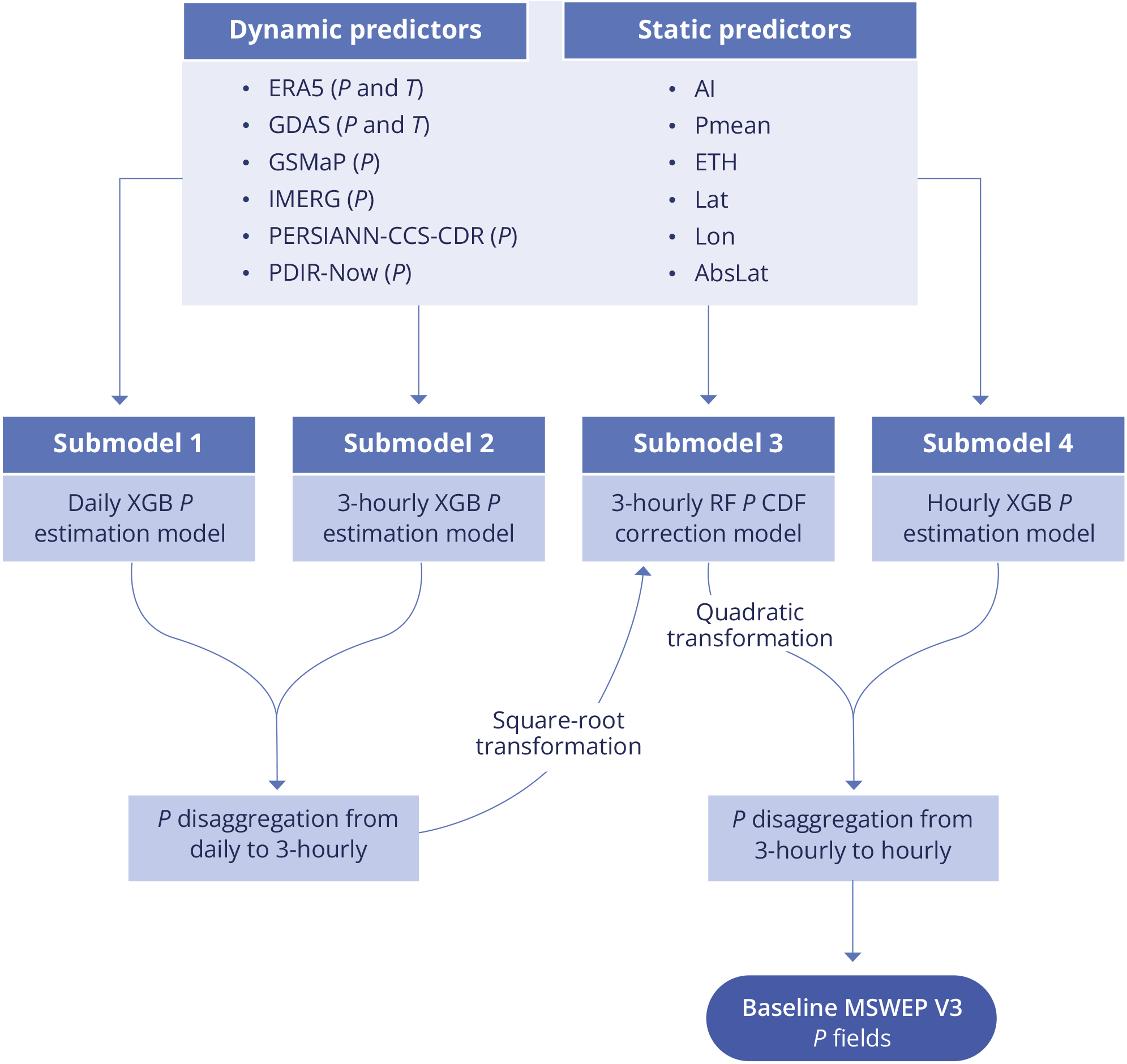} 
\caption{Flowchart of the ML model stack used to produce the hourly baseline MSWEP~V3 $P$ fields.}
\label{Flowchart}
\end{figure}

\begin{figure}
\centering
\includegraphics[width=0.93\linewidth]{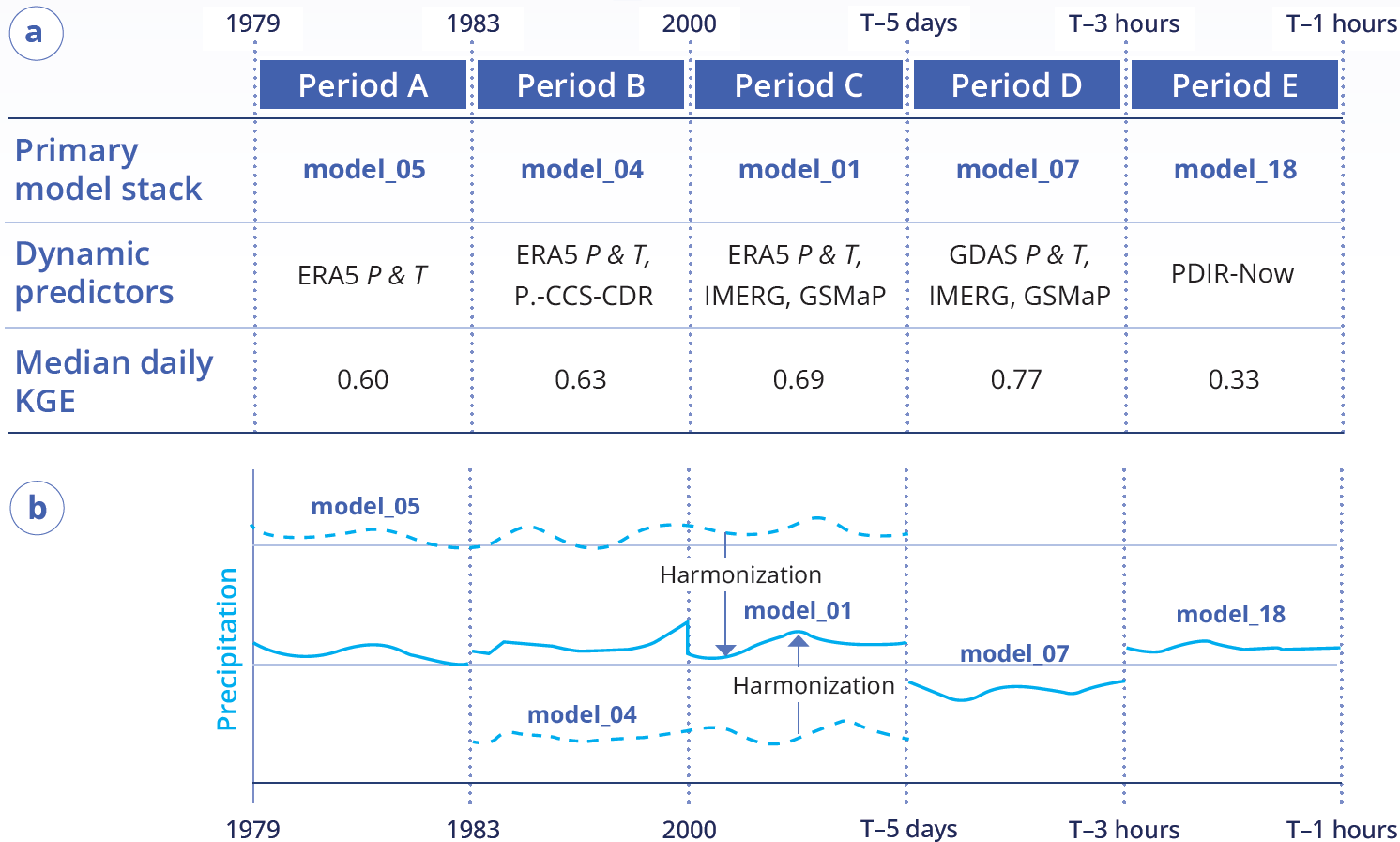} 
\caption{(a)~Different ML model stacks were used for different periods and locations to account for spatiotemporal differences in the availability of dynamic predictors. The primary ML model stacks, associated dynamic predictors, and mean daily independent validation Kling-Gupta Efficiency (KGE; from Table\ref{summary_performance_table_Global}) are also shown. Note that alternative model stacks may be applied for a given period if any dynamic predictor is unavailable. (b)~Conceptual illustration of how $P$ estimates from different ML model stacks are combined and harmonized with the reference (model\_01) in the historical production pipeline.}
\label{Periods}
\end{figure}

\begin{sidewaystable}[h!]
    \centering
    \footnotesize
    \caption{The dynamic predictors incorporated in 18 ML model stacks.}
\begin{tabular}{lrrrrrrrr}
\toprule
 & ERA5 $P$ & ERA5 $T$ & GDAS $P$ & GDAS $T$ & IMERG-L~V07 & GSMaP-MVK~V8 & P-CCS-CDR & PDIR-Now \\
\midrule
01 & \cmark & \cmark & \xmark & \xmark & \cmark & \cmark & \xmark & \xmark \\
02 & \cmark & \cmark & \xmark & \xmark & \cmark & \xmark & \xmark & \xmark \\
03 & \cmark & \cmark & \xmark & \xmark & \xmark & \cmark & \xmark & \xmark \\
04 & \cmark & \cmark & \xmark & \xmark & \xmark & \xmark & \cmark & \xmark \\
05 & \cmark & \cmark & \xmark & \xmark & \xmark & \xmark & \xmark & \xmark \\
06 & \cmark & \cmark & \cmark & \cmark & \cmark & \cmark & \xmark & \xmark \\
07 & \xmark & \xmark & \cmark & \cmark & \cmark & \cmark & \xmark & \xmark \\
08 & \cmark & \cmark & \cmark & \cmark & \cmark & \xmark & \xmark & \xmark \\
09 & \cmark & \cmark & \cmark & \cmark & \xmark & \cmark & \xmark & \xmark \\
10 & \xmark & \xmark & \cmark & \cmark & \cmark & \xmark & \xmark & \xmark \\
11 & \xmark & \xmark & \cmark & \cmark & \xmark & \cmark & \xmark & \xmark \\
12 & \xmark & \xmark & \xmark & \xmark & \cmark & \cmark & \xmark & \xmark \\
13 & \xmark & \xmark & \xmark & \xmark & \cmark & \xmark & \xmark & \xmark \\
14 & \xmark & \xmark & \xmark & \xmark & \xmark & \cmark & \xmark & \xmark \\
15 & \xmark & \xmark & \cmark & \cmark & \xmark & \xmark & \xmark & \cmark \\
16 & \cmark & \cmark & \cmark & \cmark & \xmark & \xmark & \xmark & \xmark \\
17 & \xmark & \xmark & \cmark & \cmark & \xmark & \xmark & \xmark & \xmark \\
18 & \xmark & \xmark & \xmark & \xmark & \xmark & \xmark & \xmark & \cmark \\
\bottomrule
\end{tabular}

        \label{model_predictors_table}
\end{sidewaystable}

\subsection{Assessment of Baseline Precipitation}
\label{methodsbaselinevalidation}

We used the 15,958 randomly selected, fully independent evaluation gauges (see Appendix~A ``\nameref{DataPrecipGaugeObs}'') to carry out a rigorous performance assessment of the ML model stacks underpinning the MSWEP~V3 baseline, as well as other $P$ products. We used a wide range of daily and monthly performance metrics to evaluate different critical aspects of the $P$ time series. Appendix~A ``\nameref{validationmetrics}'' provides details of each metric. 


\subsection{Assessment of Gauge Correction Scheme}
\label{assessment_gauge_corr_scheme}

To evaluate the daily gauge correction procedure, leave-one-out cross-validation was used. Each of the 57,666 daily gauges that passed the quality control (see Appendix~A ``\nameref{DailyGaugeDataQC}'') were withheld in turn and treated as an ungauged location, while the correction was performed using surrounding daily gauges with at least 70\% data overlap. The difference in performance between the uncorrected and corrected baselines was calculated using the withheld gauge as reference. Daily Pearson correlation was used as a performance metric, as the correction targets only daily variability. 





\section{Results and Discussion}
\label{ResultsandDiscussion}

\subsection{Validation of Baseline Precipitation}
\label{baselinevalidationresults}

\begin{table}[h!]
\centering
\scriptsize 
\footnotesize
\caption{Performance of the ML models underlying MSWEP~V3 (models\_01--18; prior to the harmonization step) and other state-of-the-art $P$ products, sorted by decreasing median KGE. Values represent medians across all independent evaluation gauges worldwide. Because the $P$ products span different time periods, the evaluation data varies among products. Units: $B_\textrm{peak}$ in~\% and $B_\textrm{wet days}$ in days. N$_\textrm{obs}$ indicates the number of gauges used to compute each median score.}
\centerline{\setlength{\tabcolsep}{4pt}
\begin{tabular}{rrrrrrrrrrrrrrl}
\toprule
                       & KGE & $r_\textrm{dly}$ & $\beta$ & $|\beta-1|$ & $\gamma$ & $|\gamma-1|$ & $r_\textrm{mon}$ & $B_\textrm{peak}$ & $|B_\textrm{peak}|$ & $B_\textrm{wet\,days}$ & $|B_\textrm{wet\,days}|$ & CSI$_\textrm{10\,mm}$ & N$_\textrm{obs}$ \\
\midrule
               model\_06 & 0.78 &   0.84 &    0.98 &       0.09 &     1.05 &         0.07 &            0.90 &             1.74 &              13.20 &                 -9.74 &                   12.53 &                  0.41 &             9370 \\
               model\_07 & 0.77 &            0.82 &    0.97 &       0.09 &     1.03 &         0.07 &            0.89 &             0.00 &              13.44 &                 -6.67 &                   10.16 &                  0.39 &             9370 \\
               model\_08 & 0.77 &            0.83 &    0.97 &        0.09 &     1.06 &         0.08 &            0.89 &             1.61 &              13.75 &                -11.17 &                   14.10 &                  0.44 &            10882 \\
               model\_09 & 0.76 &            0.82 &    0.97 &        0.09 &    1.05 &         0.08 &            0.89 &             1.65 &              13.73 &                -10.10 &                   12.98 &                  0.39 &             9370 \\
               model\_10 & 0.76 &            0.82 &    0.97 &       0.09 &     1.03 &         0.07 &            0.88 &             0.00 &              13.82 &                 -7.59 &                   11.90 &                  0.42 &            10882 \\
               model\_11 & 0.75 &            0.80 &    0.98 &        0.09 &     1.03 &         0.07 &            0.87 &             0.00 &              13.75 &                 -7.75 &                   11.55 &                  0.37 &             9370 \\
               model\_16 & 0.70 &            0.78 &    0.98 &        0.09 &     1.09 &         0.11 &            0.88 &             7.74 &              16.54 &                -14.89 &                   17.72 &                  0.39 &            10882 \\
               model\_15 & 0.70 &            0.76 &    0.97 &        0.10 &     1.04 &         0.08 &            0.86 &             0.04 &              14.76 &                 -8.58 &                   12.78 &                  0.33 &             9338 \\
               model\_02 & 0.69 &            0.76 &    0.97 &        0.09 &     1.05 &         0.08 &            0.85 &             2.50 &              12.43 &                 -8.72 &                   14.19 &                  0.44 &            14226 \\
               model\_01 & 0.69 &            0.76 &    0.98 &        0.09 &     1.04 &         0.07 &            0.86 &             2.10 &              11.24 &                 -6.74 &                   12.76 &                  0.42 &            12688 \\
               model\_03 & 0.68 &            0.75 &    0.98 &        0.09 &     1.04 &         0.07 &            0.86 &             1.63 &              11.72 &                 -6.56 &                   13.09 &                  0.40 &            12688 \\
           MSWEP V2.8 & 0.67 &            0.76 &    0.97 &        0.10 &     0.85 &         0.15 &            0.86 &           -20.61 &              21.55 &                 23.99 &                   24.89 &                  0.40 &            14218 \\
               model\_17 & 0.67 &            0.73 &    0.97 &        0.10 &     1.06 &         0.09 &            0.84 &             2.72 &              16.28 &                -11.17 &                   14.59 &                  0.34 &            10882 \\
                 GDAS & 0.65 &            0.73 &    1.03 &        0.14 &     0.89 &         0.12 &            0.84 &            -9.44 &              17.66 &                 17.82 &                   20.93 &                  0.34 &            10882 \\
               model\_12 & 0.63 &            0.69 &    0.98 &        0.09 &     0.98 &         0.06 &            0.75 &            -3.59 &              11.95 &                  1.85 &                   11.34 &                  0.34 &            12688 \\
               model\_04 & 0.63 &            0.70 &    0.97 &        0.10 &     1.07 &         0.10 &            0.85 &             4.65 &              14.80 &                 -9.51 &                   15.62 &                  0.36 &            12682 \\
      IMERG-Final V7 & 0.62 &            0.71 &    1.02 &        0.11 &     0.99 &         0.14 &            0.88 &             2.50 &              13.30 &                  2.89 &                   28.30 &                  0.38 &            14226 \\
               model\_13 & 0.61 &            0.66 &    0.97 &       0.09  &     0.98 &         0.07 &            0.71 &            -4.50 &              13.04 &                  3.54 &                   12.70 &                  0.34 &            14226 \\
                 ERA5 & 0.61 &            0.70 &    1.03 &        0.11 &     0.84 &         0.16 &            0.84 &           -17.47 &              19.05 &                 25.55 &                   25.90 &                  0.37 &            14226 \\
               model\_05 & 0.60 &            0.67 &    0.98 &        0.10 &     1.09 &         0.12 &            0.82 &             8.71 &              18.28 &                -12.99 &                   17.11 &                  0.36 &            14226 \\
               model\_14 & 0.58 &            0.64 &    0.97 &        0.10 &     0.96 &         0.07 &            0.72 &            -6.09 &              13.00 &                 -1.16 &                   17.13 &                  0.31 &            12687 \\
                JRA-3Q & 0.57 &            0.67 &    1.06 &        0.14 &     0.84 &         0.16 &            0.81 &           -16.33 &              18.77 &                 16.43 &                   19.85 &                  0.35 &            14226 \\
       CPC Unified & 0.54 &            0.63 &    0.88 &        0.18 &     0.95 &         0.11 &            0.83 &           -18.18 &              19.61 &                  3.22 &                   23.92 &                  0.30 &            13754 \\
      SM2RAIN-SMOS & 0.52 &            0.67 &    0.95 &        0.13 &     0.72 &         0.28 &            0.76 &           -35.94 &              36.36 &                 31.60 &                   33.08 &                  0.31 &            11940 \\
     IMERG-Early V7 & 0.47 &            0.65 &    0.88 &        0.23 &     1.03 &         0.19 &            0.71 &            -5.70 &              18.44 &                  2.58 &                   34.28 &                  0.31 &            14226 \\
     IMERG-Late V7 & 0.46 &            0.67 &    0.86 &        0.24 &     1.08 &         0.20 &            0.71 &            -3.45 &              18.24 &                 -7.05 &                   32.17 &                  0.32 &            14226 \\
         GSMaP V8 & 0.38 &            0.61 &    0.99 &        0.21 &     1.27 &         0.29 &            0.69 &            24.36 &              31.34 &                -12.16 &                   34.80 &                  0.30 &            12687 \\
          CHIRPS V2.0 & 0.36 &            0.46 &    0.99 &        0.12 &     0.94 &         0.19 &            0.85 &            -6.16 &              22.25 &                -21.01 &                   26.66 &                  0.19 &             8791 \\
           CMORPH RAW & 0.33 &            0.62 &    0.64 &        0.42 &     1.23 &         0.26 &            0.64 &           -17.46 &              25.24 &                -30.30 &                   39.55 &                  0.19 &             9365 \\
    PDIR-Now & 0.33 &            0.39 &    1.06 &        0.15 &     0.94 &         0.13 &            0.59 &            -1.92 &              18.90 &                 22.54 &                   24.67 &                  0.17 &            12647 \\
        SM2RAIN-ASCAT & 0.33 &            0.53 &    0.96 &        0.12 &     0.57 &         0.43 &            0.68 &           -47.99 &              48.20 &                 78.79 &                   78.91 &                  0.22 &            12253 \\
               model\_18 & 0.33 &            0.39 &    0.96 &        0.13 &     0.87 &         0.14 &            0.59 &           -20.80 &              24.67 &                 10.64 &                   13.83 &                  0.14 &            12647 \\
                CHIRP & 0.31 &            0.47 &    0.99 &        0.15 &     0.60 &         0.40 &            0.66 &           -41.59 &              41.77 &                 77.99 &                   78.01 &                  0.13 &             8984 \\
          SM2RAIN-CCI & 0.30 &            0.47 &    1.03 &        0.15 &     0.63 &         0.37 &            0.70 &           -42.40 &              43.41 &                 59.14 &                   59.56 &                  0.19 &            11115 \\
            CMORPH-RT & 0.28 &            0.55 &    0.89 &        0.21 &     1.38 &         0.40 &            0.59 &            27.20 &              38.51 &                -28.34 &                   37.20 &                  0.23 &             9443 \\
      PERSIANN-CCS-CDR & 0.24 &            0.35 &    1.11 &        0.19 &     1.09 &         0.23 &            0.65 &            24.77 &              28.60 &                -15.05 &                   38.42 &                  0.17 &            12665 \\
         PERSIANN-CCS & 0.17 &            0.30 &    0.88 &        0.30 &     1.09 &         0.25 &            0.42 &             2.63 &              18.97 &                -15.50 &                   42.05 &                  0.12 &            12679 \\
\bottomrule
\end{tabular}

}
    \label{summary_performance_table_Global}
\end{table}

\begin{table}[h!]
\centering
\scriptsize 
\footnotesize
\caption{Median daily KGE values for the different $P$ products across the five major K\"{o}ppen-Geiger climate classes, computed using the 15,958 independent evaluation gauges. In each column, the best-performing product is shown in bold. The gauges were classified using the 1-km resolution K\"{o}ppen-Geiger map for 1991--2020 from \citet{beck_high-resolution_2023}. For MSWEP~V3, only the primary historical model stack (model\_01) and the main NRT model (model\_07) are shown. Because the $P$ products span different time periods, the evaluation data varies among products. N$_\textrm{obs}$ indicates the number of gauges used to compute each median.}
\centerline{\setlength{\tabcolsep}{4pt}

\begin{tabular}{lcccccc}
\toprule
\textbf{Dataset} & \textbf{All} & \textbf{Tropical (A)} & \textbf{Arid (B)} & \textbf{Temperate (C)} & \textbf{Cold (D)} & \textbf{Polar (E)} \\
\midrule
N$_\textrm{obs}$ & 14190 & 1635 & 2198 & 5948 & 4296 & 113 \\
\midrule
model\_07   & \textbf{0.77} & \textbf{0.56} & \textbf{0.60} & \textbf{0.81} & \textbf{0.80} & \textbf{0.69} \\
model\_01   & 0.69 & 0.38 & 0.59 & 0.77 & 0.79 & 0.50 \\
MSWEP V2.8            & 0.67 & 0.33 & 0.56 & 0.72 & 0.72 & 0.35 \\
GDAS                   & 0.65 & 0.38 & 0.52 & 0.68 & 0.67 & 0.22 \\
IMERG-Final V7        & 0.62 & 0.37 & 0.55 & 0.68 & 0.63 & 0.34 \\
ERA5                   & 0.61 & 0.26 & 0.50 & 0.66 & 0.68 & 0.21 \\
JRA-3Q                 & 0.57 & 0.15 & 0.41 & 0.65 & 0.66 & 0.21 \\
CPC Unified            & 0.54 & 0.27 & 0.48 & 0.59 & 0.54 & 0.40 \\
SM2RAIN-SMOS           & 0.52 & 0.23 & 0.42 & 0.59 & 0.58 & 0.26 \\
IMERG-Early V7         & 0.47 & 0.34 & 0.38 & 0.53 & 0.45 & 0.29 \\
IMERG-Late V7          & 0.46 & 0.36 & 0.42 & 0.50 & 0.43 & 0.29 \\
GSMaP V8               & 0.38 & 0.34 & 0.38 & 0.42 & 0.35 & 0.20 \\
CHIRPS V2.0            & 0.36 & 0.29 & 0.41 & 0.40 & 0.31 & 0.24 \\
CMORPH-RAW             & 0.33 & 0.45 & 0.45 & 0.26 & 0.36 & 0.20 \\
PDIR-Now               & 0.33 & 0.24 & 0.25 & 0.36 & 0.36 & 0.23 \\
SM2RAIN-ASCAT          & 0.33 & 0.12 & 0.29 & 0.38 & 0.39 & 0.16 \\
CHIRP                  & 0.31 & 0.17 & 0.23 & 0.36 & 0.42 & 0.23 \\
SM2RAIN-CCI            & 0.30 & 0.13 & 0.23 & 0.34 & 0.37 & -0.38 \\
CMORPH-RT              & 0.28 & 0.41 & 0.37 & 0.27 & 0.25 & 0.10 \\
PERSIANN-CCS-CDR       & 0.24 & 0.24 & 0.24 & 0.24 & 0.23 & 0.08 \\
PERSIANN-CCS           & 0.16 & 0.19 & 0.14 & 0.18 & 0.13 & -0.07 \\
\bottomrule
\end{tabular}

}
    \label{daily_KGE_KGzones}
\end{table}


MSWEP~V3 baseline performance was evaluated for 15,958 independent gauges withheld from training using a range of daily metrics sensitive to different aspects of $P$ time series (subsection~\ref{DataMethods}\ref{methodsbaselinevalidation}). Across the 18 stacks (model\_01--model\_18)---each using a different combination of dynamic predictors (Table~\ref{model_predictors_table})---the median Kling-Gupta Efficiency (KGE) ranges from 0.33 for model\_18, which relies solely on PDIR-Now and is used only for estimating NRT $P$ before IMERG and GSMaP are released, to 0.78 for model\_06, which incorporates GDAS, ERA5, IMERG, and GSMaP (Table~\ref{summary_performance_table_Global} and Fig.~\ref{Periods}). The reference stack, model\_01, incorporating three dynamic predictors (ERA5, IMERG, GSMaP) and covering the largest portion of the record (Year 2000 to 5~days before present), achieved a median KGE of 0.69, a very small peak bias ($B_{\textrm{peak}}=2.1$\%), a slight wet-day underestimation ($B_{\textrm{wet\,days}}=-6.7$~days), and strong event-detection skill (CSI$_{10\,\textrm{mm}}=0.42$). In contrast, stacks relying on just a single dynamic predictor (e.g., model\_14 and model\_05) yield relatively lower median KGE values (0.58 and 0.60, respectively), larger peak biases ($-6.1$\% and $+8.7$\%), and lower CSI$_{10,\textrm{mm}}$ (0.31 and 0.36). 


When compared with other (quasi-)global gridded $P$ products for the 15,958 independent evaluation gauges, the MSWEP~V3 baseline outperforms them across nearly all metrics (Table~\ref{summary_performance_table_Global}). The reference stack model\_01 (median daily KGE=0.69)---positioned roughly mid-range within the 18-model ranking---exceeds all other products, whose median KGE values span from 0.17 (PERSIANN-CCS) to 0.65 (GDAS). It also outperforms widely used non-gauge-corrected products in terms of KGE such as CHIRP (0.31), ERA5 (0.61), GSMaP~V8 (0.38), and IMERG-L V7 (0.46). GDAS performs best among all products other than MSWEP~V3. However, as an evolving operational atmospheric analysis, its high performance only applies to 2021 onward (after the NOAA Global Forecast System---GFS---was upgraded to V16). model\_01 also outperforms products that directly incorporate gauge observations, including CHIRPS (KGE of 0.36), CPC Unified (0.54), and IMERG-F~V7 (0.62). CHIRPS applies 5-day corrections and IMERG monthly corrections, which offer limited benefit in this daily evaluation. Finally, model\_01 outperforms the previous MSWEP version (V2.8) with daily gauge corrections (median KGE=0.67). These results underscore the value of multi-source, ML-based approaches for mitigating biases and capturing diverse $P$ dynamics worldwide (see reviews by \citealp{hussein_rainfall_2022,dotse_review_2024,papacharalampous_comparison_2023,xu_multi-source_2024}).

\subsection{Precipitation Correlation Lengths}
\label{results_correlation_lengths}

\begin{figure} 
    \centering  
    \includegraphics[width=0.8\linewidth]{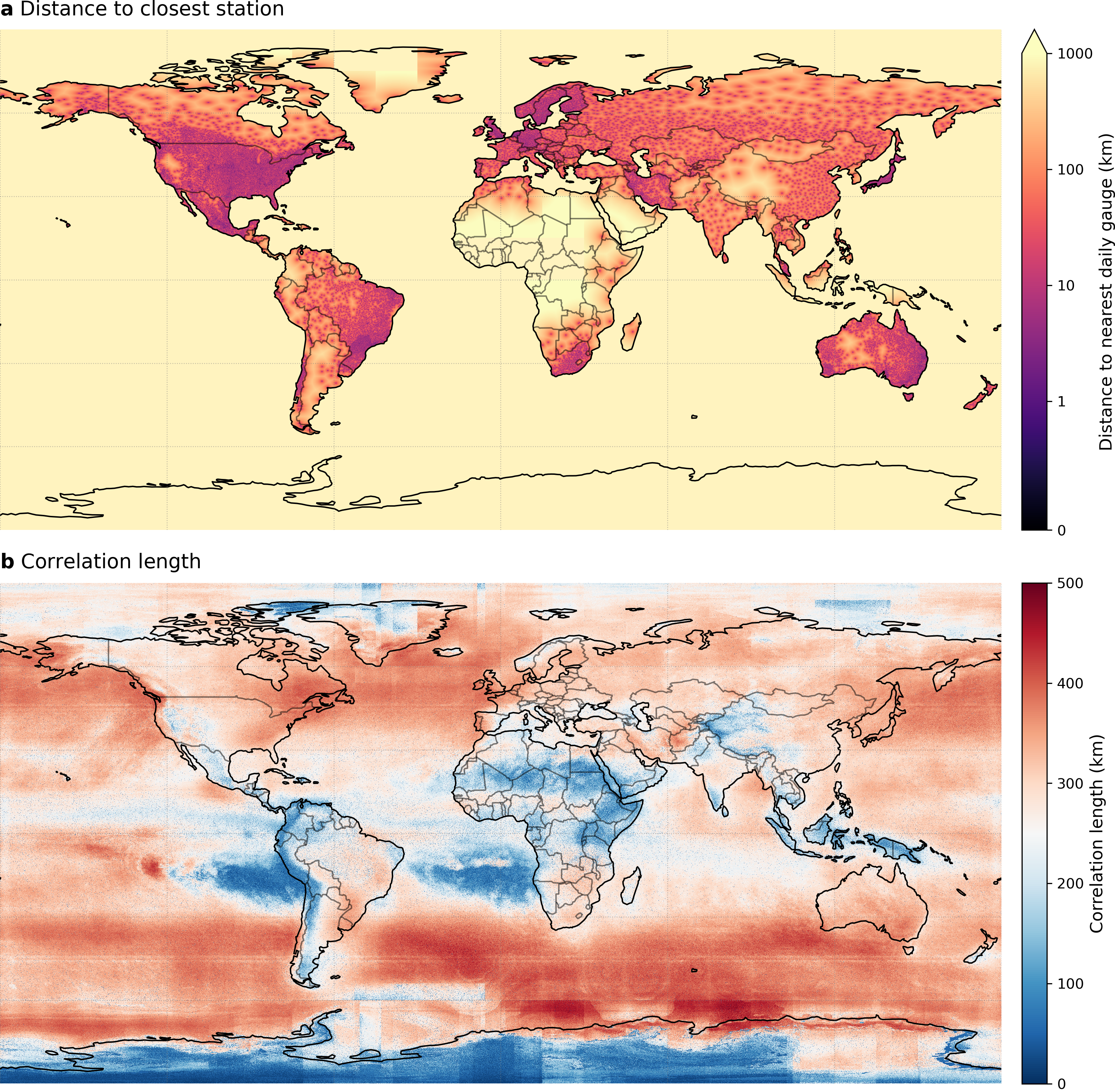} 
    \caption{(a)~Distance to the closest daily gauge that passed quality control and was used in the daily gauge-correction procedure ($n=57,666$; Appendix~A ``\nameref{DailyGaugeDataQC}''). (b)~$P$ correlation lengths based on the baseline MSWEP~V3 (Appendix~A, ``\nameref{MethodsCorrelationLengths}'').}
    \label{Daily_Gauge_CorrectionF1} 
\end{figure}

Fig.~\ref{Daily_Gauge_CorrectionF1}b shows the spatial correlation length (i.e., e-folding distance) of daily $P$ globally based on the MSWEP~V3 baseline (Appendix~A, ``\nameref{MethodsCorrelationLengths}''). These lengths summarize how heterogeneous daily $P$ fields are and control how rapidly a gauge's corrective influence decays with distance when adjusting nearby grid cells. The global mean correlation length is 289~km; over all land (without Antarctica) it is 262~km; over ocean it is 300~km. Over all land, values range from 145~km (10th percentile) to 342~km (90th percentile), with a median of 281~km. Short correlation lengths ($<200$~km)---indicative of small spatial event scales and limited spatial gauge influence---prevail in arid Africa, the southeast Asian archipelagos, and in mountain belts such as the northern Andes and the Hindu Kush, where $P$ is dominated by localized deep convection and strong orographic forcing \citep{schumacher_assessing_2023,roe_orographic_2005}. Short correlation lengths in high-latitude ice-covered regions (e.g., Greenland and Antarctica) likely reflect infrequent, isolated $P$ occurrences confined to a small number of grid cells. Long correlation lengths ($>350$~km)---reflecting large event scales and broader gauge influence---prevail in the eastern US, western Europe, along the west coast of Canada, in central Chile, eastern East Asia, and southern Australia, where midlatitude fronts and atmospheric rivers produce widespread, synoptic-scale stratiform $P$ \citep{schumacher_assessing_2023}. 


Previous studies that mapped $P$ correlation lengths include \citet{funk_climate_2015}, \citet{smith_spatial_2005}, \citet{kursinski_spatiotemporal_2008}, \citet{touma_characterizing_2018}, and \citet{fan_spatial_2021}. Our results generally align with these studies, although differences are expected because of variations in data sources, $P$ accumulation periods, $P$ thresholds, and study periods. In particular, estimates based on satellite and/or reanalysis data, such as ours based on the MSWEP~V3 baseline, may yield longer lengths because of their lower effective resolution \citep{kottek_global_2007}. Most studies computed seasonal lengths; we did not, because doing so would introduce time-varying gauge influences and compromise the homogeneity of the gauge-corrected MSWEP~V3. We also assumed isotropy (no directional dependence), despite documented anisotropy along mountain ranges and prevailing winds (e.g., \citealp{kursinski_spatiotemporal_2008,chen_satellite_2016,fan_spatial_2021}). Incorporating anisotropy may be explored in future MSWEP releases.

\subsection{Gauge Reporting Times}
\label{results_gauge_reporting_time}

\begin{figure} 
        \centering  
        \includegraphics[width=\linewidth]{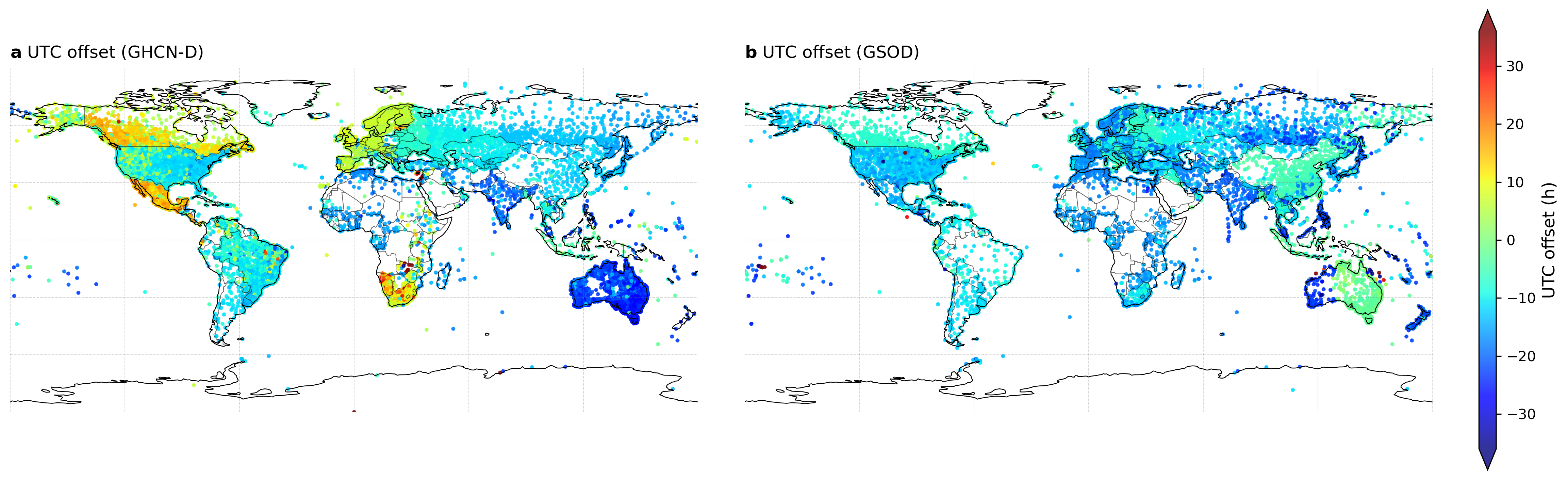} 
 \caption{Inferred reporting-time offsets (hours relative to midnight UTC) for daily gauges in (a)~GHCN-D and (b)~GSOD. For example, an offset of $-6$~h indicates that the daily total spans 18:00~UTC on the previous day to 18:00~UTC on the current day. Reporting times were inferred by maximizing agreement with the hourly MSWEP~V3 baseline (Appendix~A ``\nameref{DailyGaugeDataQC}'').}
        \label{UTC_offset} 
\end{figure}

Reporting times vary substantially across time zones, countries, and regions, depending on local observational practices \citep{yang_development_2020}. Fig.~\ref{UTC_offset} shows daily reporting-time offsets (hours relative to midnight UTC) for the GHCN-D and GSOD databases inferred from the hourly MSWEP~V3 baseline $P$ estimates. For GHCN-D (Fig.~\ref{UTC_offset}a), offsets exceed $+10$~h UTC across much of Canada, Mexico, and southern Africa; are around $+8$~h over western Europe; around $-10$~h over most of the United States, eastern Europe, east Africa, Russia, and east Asia; around $-16$~h over west Africa; around $-20$~h over India; and $<-24$~h over Australia. For GSOD (Fig.~\ref{UTC_offset}b)---automated gauges that officially report 00:00--24:00~UTC totals---the inferred offsets are all negative, ranging from about $-4$ to $-20$~h UTC depending on the region. This indicates that GSOD daily accumulations still include $P$ from the previous day, consistent with the official documentation \citep{noaa_ncep_global_2020}. 

Our reporting times are consistent with those from previous global analyses \citep{beck_mswep_2019} and with regional studies for the US \citep{belcher_method_2005}, Europe \citep{overeem_euradclim_2023}, the Middle East \citep{wang_saudi_2025}, Asia \citep{yatagai_end_2020}, and Australia \citep{contractor_how_2015}. While satellite products like IMERG and consequently MSWEP~V3 detect the start, peak, and end of $P$ events earlier (on average by 26, 17, and 7~minutes respectively)---due to ice-scattering signatures before surface $P$ and uncorrected parallax shifts \citep{li_how_2023}---these discrepancies are too small to substantially affect the inferred reporting times. 

This substantial variability in daily reporting times---both between and within countries and among data sources---makes it crucial to explicitly account for them when training ML models, applying gauge corrections, and evaluating $P$ products using daily gauge observations. To illustrate the impact of considering reporting times, using standard 00:00--24:00~UTC IMERG-L V7 daily totals yields a mean Pearson correlation of 0.41 across 15,959 training stations. When IMERG-L V7 daily totals are aligned with each station's actual reporting time, the correlation increases to 0.52. This corresponds to a substantial 49\% increase in explained variance ($100 \times \frac{0.52^2}{0.41^2} - 100$). Despite this, most studies neither report nor account for gauge reporting times. 

\subsection{Daily Gauge Correction}
\label{dailygaugecorrectionresults}

The spatial distribution of daily rain gauges is highly uneven across the globe. Fig.~\ref{Daily_Gauge_CorrectionF1}a maps the distance to the nearest daily gauge after quality control ($n=57,666$; see Appendix~A ``\nameref{DailyGaugeDataQC}''). The global median distance over all land (without Antarctica) is 78.7~km, and only 21.2\% of global land area (without Antarctica) has a daily gauge within 25~km. Short distances ($<20$~km) are found across parts of Europe, Iran, eastern Australia, the CONUS, Mexico, southeastern Brazil, South Africa, and Japan. Large distances ($>100$~km) dominate most of Africa (except South Africa), high northern latitudes ($>60^\circ$N; except Scandinavia), the Tibetan Plateau, the Arabian Peninsula, and parts of South America (Colombia, Ecuador, Peru, Bolivia, Paraguay and Argentina). These patterns reflect persistent socioeconomic and logistical disparities---dense networks in developed mid-latitudes and sparse coverage across much of the tropics, complex terrain, and deserts---leaving vast regions underrepresented for daily validation and baseline correction \citep{kidd_so_2017}. Note that this map draws primarily on global archives (GHCN-D and GSOD), so low apparent gauge density may reflect incomplete reporting to these databases rather than sparse national networks.

\begin{figure} 
        \centering  
        \includegraphics[width=1\linewidth]{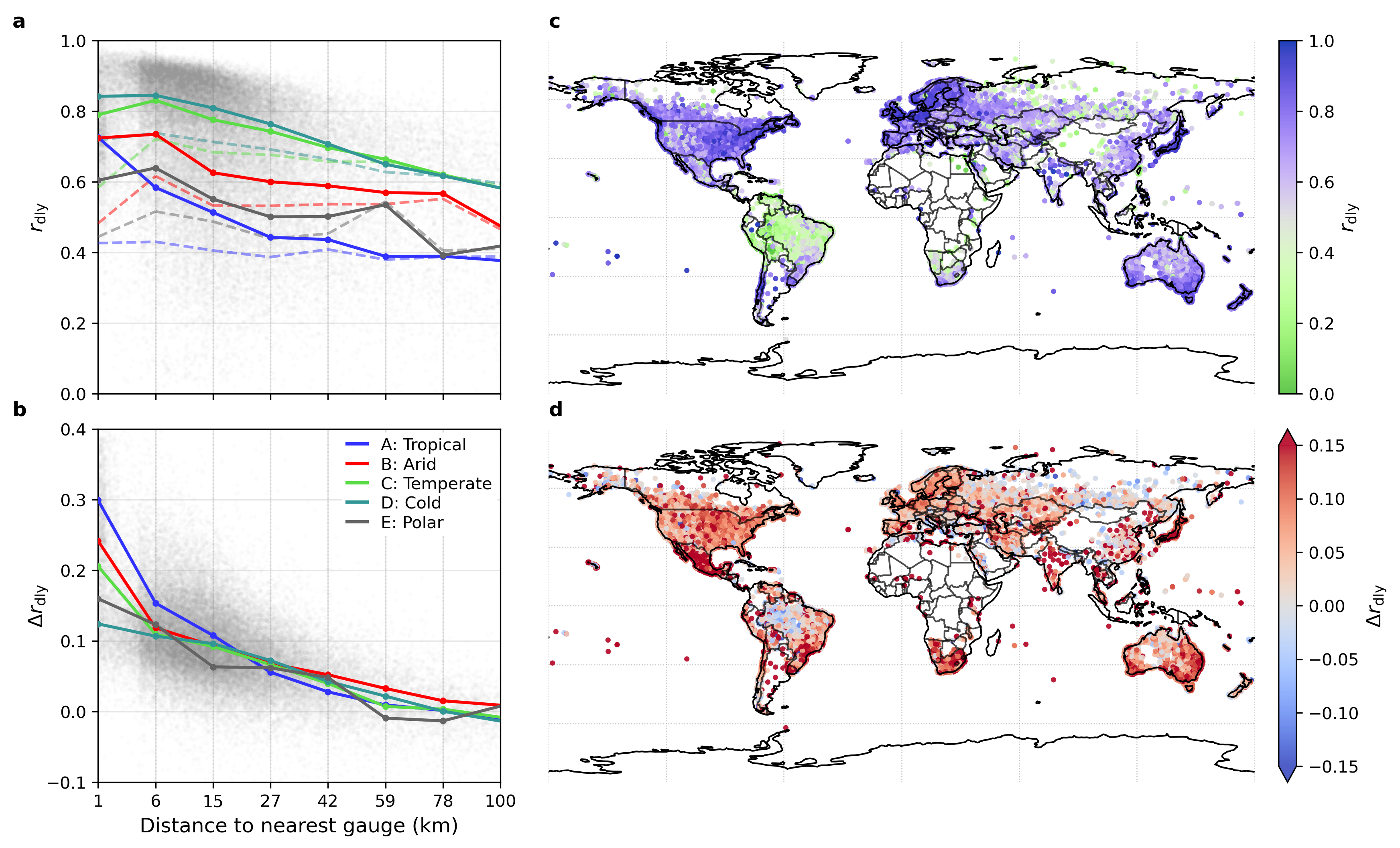} 
        \caption{(a)~Daily temporal correlation ($r_\textrm{dly}$) after gauge correction of the baseline versus distance to the nearest gauge. Each gray point ($n=57,666$) denotes a gauge for which post-correction $r_\textrm{dly}$ was evaluated using surrounding gauges (Appendix~A ``\nameref{MethodsDailyGaugeCorrection}''). Solid lines show segmented regression fits to the post-correction $r_\textrm{dly}$ for each major K\"{o}ppen-Geiger class \citep{beck_high-resolution_2023}; dashed lines show fits to the baseline (uncorrected) $r_\textrm{dly}$. The $x$-axis tick labels indicate the breakpoints used in the segmented regressions. (b)~Change in $r_\textrm{dly}$ after gauge correction ($\Delta r_\textrm{dly}$) versus distance to the nearest gauge. Each gray point ($n=57,666$) is the difference between post-correction and baseline $r_\textrm{dly}$; solid lines show segmented regression fits for each K\"{o}ppen-Geiger class. (c)~Global distribution of post-correction daily temporal correlation ($r_\textrm{dly}$) evaluated at gauge locations. (d)~Global distribution of the change in daily temporal correlation caused by gauge correction ($\Delta r_\textrm{dly}$).}
        \label{deltaR_figure} 
\end{figure}

We evaluated the effectiveness of the OI-based daily gauge-correction scheme using leave-one-out cross-validation. For each gauge ($n=57{,}666$), we withheld its observations, treated the location as ungauged, and computed the change in daily temporal correlation ($\Delta r_\textrm{dly}$; Fig.~\ref{deltaR_figure}d; subsection~\ref{DataMethods}\ref{assessment_gauge_corr_scheme}). We focus on $r_\textrm{dly}$ because the daily gauge correction procedure is designed to improve daily variability (Appendix~A ``\nameref{MethodsDailyGaugeCorrection}''). The median $\Delta r_\textrm{dly}$ is +0.09 (10th percentile: +0.02; 90th percentile: +0.20). Despite relatively short distances to the nearest available gauge (median distance: 14~km; mean: 19~km), average improvements remain modest due to the already strong baseline performance (median $r_\textrm{dly} = 0.67$). For many gauges the improvement is negligible; this primarily reflects insufficient nearby gauges rather than a failure of the method. The largest improvements occur in regions with dense networks---Europe, the United States, Australia, and southeastern Brazil. If reporting times were ignored and daily accumulations were assumed to represent 00:00–24:00~UTC, the apparent improvement from gauge correction would be substantially larger because baseline performance would appear much worse. This further underscores the importance of explicitly accounting for gauge reporting-time offsets (subsection~\ref{ResultsandDiscussion}\ref{results_gauge_reporting_time}).

Other studies that used daily gauge observations to correct baseline $P$ fields focused on a single catchment in China \citet{hu_satellite_2015}, Brazil as a whole \citet{rozante_combining_2010}, Chile \citet{baez-villanueva_rf-mep_2020}, and China \citet{lei_two-step_2022}. These studies relied on daily (rather than hourly) baseline products—mostly satellite-based—and did not explicitly account for reporting-time offsets. Moreover, only \citet{baez-villanueva_rf-mep_2020} released a corresponding downloadable dataset.


To explore how gauge proximity influences performance improvement, we relate baseline $r_\textrm{dly}$, gauge-corrected $r_\textrm{dly}$, and their difference ($\Delta r_\textrm{dly}$) to the distance to the nearest gauge, and fit segmented regressions for the five major K\"{o}ppen-Geiger climate classes (A--E; Fig.~\ref{deltaR_figure}a,b). As expected, improvements increase as gauge distance decreases. At very short distances ($<10$~km), the largest gains occur in tropical (A) climates (median $\Delta r_\textrm{dly}$ of +0.22), while the smallest improvements occur in cold (D, +0.10) climates; the other classes fall in between. This mirrors baseline skill: the baseline $r_\textrm{dly}$ is lowest in the tropics (median 0.41), leaving more room for improvement, and highest in cold and temperate climates (median 0.74), leaving less opportunity for improvement. At larger distances ($>20$~km), the climatic contrast in gains largely vanishes, likely because correlation lengths are shorter in the tropics (median 245~km) than in cold regions (median 315~km; Fig.~\ref{Daily_Gauge_CorrectionF1}b). Across all climate zones, the baseline performance is slightly higher at shorter gauge distances. While this may seem counterintuitive, it is likely because shorter distances occur in well-monitored regions where more observations are assimilated into ERA5, yielding stronger baseline skill. To our knowledge, no prior study has explicitly evaluated changes in post-correction performance as a function of gauge distance, although some studies have assessed performance changes by varying gauge density (e.g., \citealp{rozante_combining_2010,hu_satellite_2015,baez-villanueva_rf-mep_2020}).


Similar to MSWEP~V3, several other gridded $P$ datasets also correct hourly baseline $P$ fields using daily gauge observations, such as PERSIANN-CCSA \citep{boushaki_bias_2009}, EURADCLIM \citep{overeem_euradclim_2023}, GSMaP\_Gauge \citep{mega_gauge_2014}, and MERRA-2 \citep{gelaro_modern-era_2017,reichle_land_2017}. However, the correction methods used in these datasets can be improved upon. GSMaP\_Gauge and MERRA-2 do not incorporate gauge distance-dependent weighting, which degrades performance in regions with sparse gauge coverage. GSMaP\_Gauge, PERSIANN-CCSA, and MERRA-2 neglect $P$ correlation length scales (see subsection~\ref{ResultsandDiscussion}\ref{results_correlation_lengths}) entirely, while EURADCLIM applies a single, fixed correlation length, potentially resulting in suboptimal corrections. Furthermore, PERSIANN-CCSA and GSMaP\_Gauge do not explicitly account for gauge reporting times, leading to temporal mismatches between daily baseline and gauge $P$ totals (see subsection~\ref{ResultsandDiscussion}\ref{results_gauge_reporting_time}). 


\subsection{Global $P$ Climatology and Trends}
\label{globalPtrends}

 \begin{figure} 
        \centering  
        \includegraphics[width=0.88\linewidth]{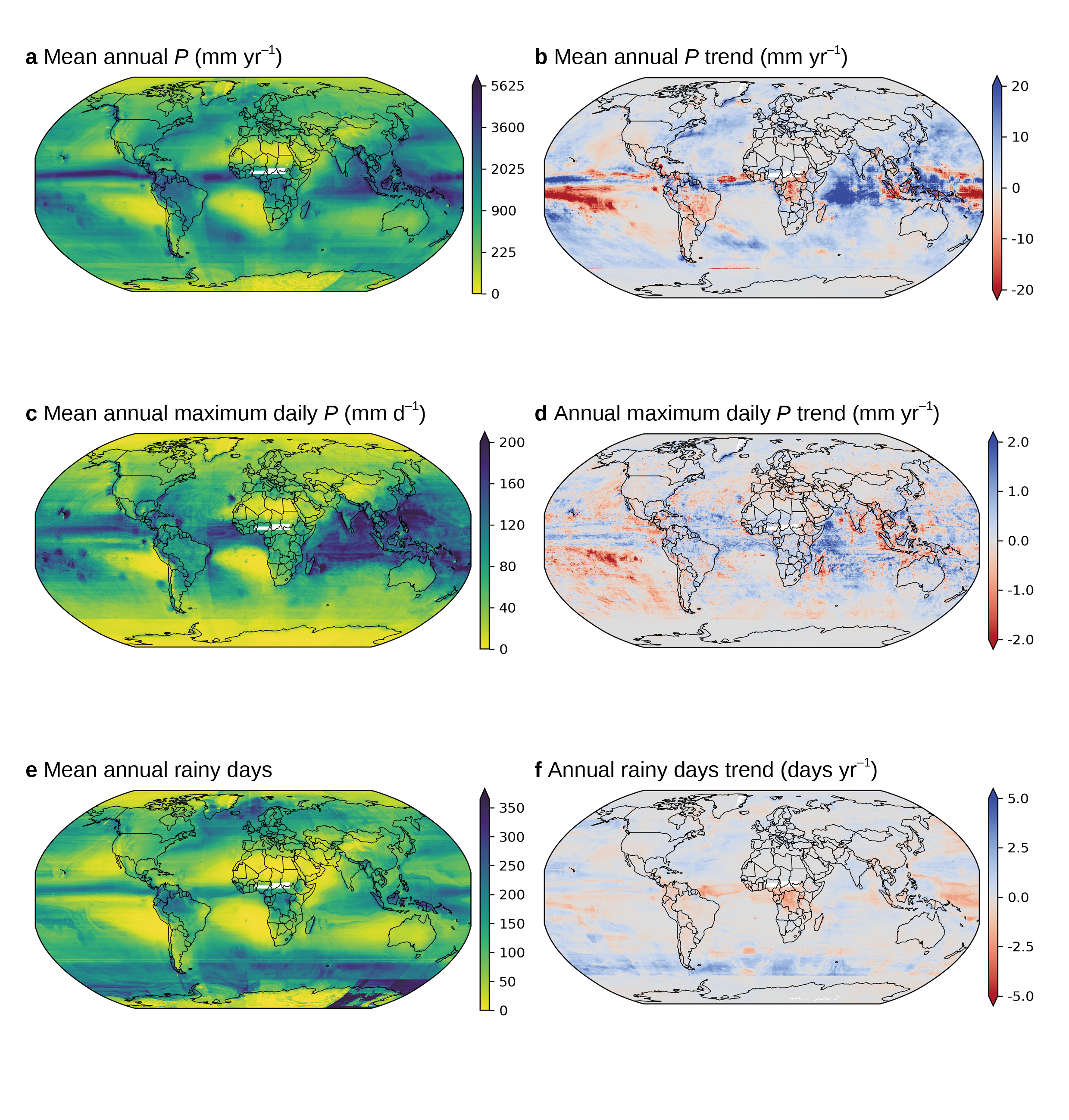} 
        \caption{Mean and trend values (1979--2024) based on the MSWEP~V3 $P$ product for (a--b)~mean annual $P$, (c--d)~annual maximum daily $P$, and (e--f)~annual number of wet days (using a threshold of 0.5~mm~d$^{-1}$). Note that panel~(a) uses a non-linear color scale.}
        \label{Global_Clim_Trends_figure} 
\end{figure}

Fig.~\ref{Global_Clim_Trends_figure} presents global climatologies and trends (1979--2020) of mean annual $P$, annual maximum daily $P$, and annual wet-day counts derived from the final, gauge-corrected MSWEP~V3 product. The mean annual $P$ over all land (without Antarctica) is 825~mm~yr$^{-1}$ (Fig.~\ref{Global_Clim_Trends_figure}a), comparable to the GPCC~V2015 estimate of 792~mm~yr$^{-1}$ \citep{beck_bias_2020}. This agreement is expected because MSWEP~V3 is bias-corrected using GPCC (Appendix~A ``\nameref{MethodsMonthlyGaugeCorrection}''). Both estimates are likely conservative \citep{beck_bias_2020}, given that (i)~GPCC relies on rain gauges, which are disproportionately located at low elevations in mountainous regions \citep{briggs_topographic_1996,schneider_gpccs_2014,kidd_so_2017}, and (ii)~gauges can underestimate $P$ by up to 90\% due to wind-induced undercatch \citep{groisman_accuracy_1994,sevruk_wmo_2009,rasmussen_how_2012}. Overall, the spatial patterns agree with other gauge-based assessments of mean $P$ \citep{funk_global_2015,fick_worldclim_2017,karger_climatologies_2017}, extreme $P$ \citep[e.g.,][]{alexander_intercomparison_2020,dunn_comparing_2022}, and $P$ frequency \citep[e.g.,][]{sun_how_2006,dietzsch_global_2017}. The mean annual maximum daily $P$ over all land (without Antarctica) in MSWEP~V3 is 58~mm~d$^{-1}$ (Fig.~\ref{Global_Clim_Trends_figure}c), higher than the 46~mm~d$^{-1}$ estimate from the global gauge- and ML-based PPDIST dataset (V1.0; \citealp{beck_ppdist_2020}), mainly due to higher tropical maxima. The mean annual number of wet days ($P\ge 0.5$~mm~d$^{-1}$) over all land (without Antarctica) is 94~days (Fig.~\ref{Global_Clim_Trends_figure}e), slightly higher than the PPDIST estimate of 80~days, again primarily due to higher tropical rain frequencies. Differences from PPDIST likely arise from: (i)~different satellite inputs (PPDIST used IMERG~V06; MSWEP V3 uses IMERG~V7 and GSMaP~V8), and (ii)~overestimation of temporal autocorrelation, resulting from blending multiple $P$ datasets, which inflates daily $P$ amounts.

The long-term trend maps (1979--2024) reveal a complex pattern of changes in $P$ characteristics (Fig.~\ref{Global_Clim_Trends_figure}b,d,f). The mean annual $P$ trend over all land (without Antarctica) is +0.13~mm~yr$^{-2}$, indicating no substantial change. In contrast, mean annual maximum daily $P$ increased by +0.07~mm~d$^{-1}$~yr$^{-1}$, and the mean number of wet days decreased by --0.03~days per year, consistent with intensifying moisture transport \citep{trenberth_changes_2011,giorgi_response_2019}. Distinct regional trend patterns are evident, largely in agreement with other global trend assessments \citep[e.g.,][]{adler_global_2018,sun_global_2021,gu_observed_2023}. Trends in annual maximum daily $P$, and in mean annual $P$ over arid regions, should be interpreted with caution because of substantial interannual variability and significant observational uncertainty across gauge, satellite, and reanalysis inputs \citep{beck_global-scale_2017,beck_daily_2019,abbas_comprehensive_2025,wang_saudi_2025}. For example, the pronounced negative trend in Central Africa reflects changes in the observing system assimilated by ERA5 \citep{hersbach_era5_2020,gleixner_did_2020}. Over the oceans and at high latitudes, artifacts arise because the lack of gauges forces the ML models to extrapolate land-based relationships into regions with no observational constraints, potentially generating spurious spatial patterns. 

\section{Conclusion}
\label{conclusions}

MSWEP~V3 represents the next generation in the MSWEP series evolution, delivering a global, hourly 0.1$^\circ$ $P$ record from 1979 to the present with a $\sim$2-hour latency. Baseline $P$ estimates are produced using ML model stacks incorporating satellite and (re)analysis $P$ and $T$ predictors along with static geographic, topographic, and climatic predictors. These baseline estimates are corrected using daily and monthly gauge data through an OI scheme that accounts for gauge proximity, reporting times, inter-gauge dependence, and spatial correlation structure. The following points summarize the main conclusions, each corresponding to a specific subsection of the Results and Discussion:
 
\begin{enumerate}

\item Hourly MSWEP~V3 baseline $P$ fields were generated using 18 distinct ML model stacks and evaluated against 15,958 independent gauges withheld from training. The reference stack, model\_01, which covers the period from 2000 to $\sim$5~days from the present, achieved a median Kling-Gupta Efficiency (KGE) of 0.69, displaying strong event-detection skill and a very small peak bias. Critically, it outperformed all 18 other gridded $P$ products assessed, including non-gauge-corrected products (e.g., CHIRP, ERA5, GSMaP V8, and IMERG-L V7) and even those that directly incorporate gauge observations (e.g., CHIRPS, CPC Unified, and IMERG-F~V7), as well as the preceding MSWEP~V2.8. These findings underscore the effectiveness of MSWEP~V3's multi-source, ML-based approach in accurately capturing diverse global $P$ dynamics and mitigating biases and errors. Despite this strong performance, further regional validation using independent rain-gauge observations and complementary evaluation through hydrological modeling remain essential.

\item We computed global daily $P$ correlation lengths from the baseline $P$ fields as  necessary inputs for the OI-based gauge correction. Correlation lengths are shorter in arid Africa, the Southeast Asian archipelagos, and mountain belts, regions characterized by localized deep convection and/or orographic forcing. In contrast, correlation lengths are longer across the eastern US, western Europe, the west coast of Canada, central Chile, East Asia, and southern Australia, regions characterized by synoptic systems and atmospheric rivers producing broad $P$ fields. Over oceans, lengths are shortest along the eastern subtropical basin flanks and within the ITCZ, and longest poleward of $\sim$30$^\circ$.

\item We inferred reporting-time offsets (in hours relative to midnight UTC) for all daily gauges worldwide by identifying the time shift that maximized correlation between shifted hourly baseline $P$ data and gauge observations. Offsets in the GHCN-D database range from strongly positive (Canada, Mexico, southern Africa) to strongly negative (US, eastern Europe, Russia, East Asia, and particularly India and Australia), while GSOD gauges---which officially report 00:00--24:00~UTC---have uniformly negative offsets (about --4 to --20~h), implying that their totals include $P$ from the previous day. Accounting for these reporting times is essential when training ML models, applying corrections, and evaluating gauge-based products, yet other studies rarely consider them.

\item Global gauge coverage is highly uneven---dense in developed mid-latitudes, sparse across Africa, high latitudes, complex terrain, and deserts. Leave-one-out tests show the OI-based gauge correction scheme increases daily temporal correlation by a median $\Delta r_\textrm{dly}$ of +0.09 (10th and 90th percentiles: +0.02 and +0.20) over the MSWEP~V3 baseline (median $r_\textrm{dly}$=0.67), with the largest gains in densely gauged regions and negligible change where gauges are distant. The improvement depends not only on gauge density but also on climate zone. $\Delta r_\textrm{dly}$ at small gauge distances ($<10$~km) is greatest in the tropics (+0.23) and smallest in cold climates (+0.10), due to shorter correlation lengths and lower baseline performance in the tropics compared to cold and temperate regions. Improvements converge among climate types beyond $\sim$20~km distance. 

\item MSWEP~V3 reproduces established global patterns in mean, extreme, and frequency-based $P$, with higher tropical extremes and wet-day counts than PPDIST, likely due to updated satellite inputs and increased temporal autocorrelation from blending multiple datasets. Trend analyses over 1979--2024 show minimal global change in mean annual $P$, but modest increases in annual maximum daily $P$ and slight declines in wet-day frequency, consistent with intensifying atmospheric moisture transport. Regional trend patterns broadly agree with previous global assessments, though substantial observational uncertainty remains, particularly in arid regions and/or areas with sparse gauge coverage.


\end{enumerate}

Overall, MSWEP~V3 provides researchers, professionals, and policymakers with information needed to tackle several pressing environmental and socio-economic challenges globally and regionally. The product delivers a high-resolution, historical and NRT resource designed to support a diverse range of applications, including water resource management, hydrological modeling, agricultural planning, detection of hydrological changes and extremes, disaster risk reduction, and climate studies. MSWEP~V3 is available at \url{www.gloh2o.org/mswep}. 


\clearpage

\appendix[A] 

\section*{Predictors}
\label{DataPredictors}

To produce the MSWEP~V3 baseline $P$ fields, we used a range of gridded satellite and (re)analysis products as dynamic predictors, where ``dynamic'' refers to time-varying inputs (Table~\ref{globalPProductDetails}). Predictors were restricted to non–gauge-based products---excluding, for example, IMERG-Final---to ensure that: (i)~a satellite- and (re)analysis-based baseline with globally consistent performance is obtained; (ii)~label leakage and overweighting are avoided, preventing the models from learning from inputs that incorporate the same gauges used as training targets; and (iii)~gauge correction remains a separate step in the MSWEP~V3 production pipeline (Appendix~A ``\nameref{MethodsMLalgorithm}'').

As dynamic predictors we used the microwave- and infrared-based IMERG-Early and -Late V07 \citep{huffman_gpm_2019} and GSMaP-MVK and -NRT~V8 \citep{kubota_global_2020} $P$ products, which cover the period 2000 to approximately 3--4~hours before real time. For earlier periods (before 2000) and for the most recent hours (3--4 hours from real time)---when microwave data are unavailable---we relied on the infrared-based PERSIANN‐CCS‐CDR \citep{sadeghi_persiann-ccs-cdr_2021} and PDIR‐Now \citep{nguyen_persiann_2020} $P$ products. We also included $P$ and $T$ data from the ECMWF ERA5 reanalysis \citep{hersbach_era5_2020} and GDAS analysis \citep{noaa_ncep_national_2025} as predictors. The $T$ data, provided as daily mean values, allow the ML models to account for seasonal variations in the error and bias characteristics of the $P$ products. Incorporating such a diverse set of predictors strengthens model robustness across climates, geographic regions, and atmospheric conditions, leading to a more accurate and reliable representation of global $P$ patterns. All predictors with native resolutions $>0.1^\circ$ were resampled to 0.1$^\circ$ using nearest-neighbor interpolation.

We also used static predictors to generate the MSWEP~V3 baseline $P$ fields, where ``static'' refers to time-invariant (constant) variables. Six predictors were selected for their relevance to $P$ generation and product performance (Table~\ref{StatPredictor}). Two are climate indices: Aridity Index (AI), calculated as long-term mean $P$ divided by potential evaporation (PE), and Pmean, the mean annual $P$. One represents topography: Effective Terrain Height (ETH), a smoothed topographic data to represent orographic effects on $P$ patterns. The remaining three describe geographic position: latitude (Lat), longitude (Lon), and absolute latitude (AbsLat). All predictors except the geographic variables were resampled to 0.1$^\circ$ using area-averaging.

\begin{table}[h!]
\centering
\footnotesize
\caption{Overview of the static predictors used in the ML models to generate MSWEP~V3 baseline $P$ fields.}
\begin{tabularx}{\textwidth}{@{} l>{\raggedright\arraybackslash}X>{\raggedright\arraybackslash}X @{}}
\toprule 
Name (units) & Description & Data Source(s) \\
\midrule
AI ($-$) & Aridity index (AI) calculated as ratio of long-term mean $P$ to PE & Mean annual $P$ from CHELSA V2.1 (1-km resolution; \citealp{karger_climatologies_2017}) and PE from \citet[][1~km]{trabucco_global_2018} for land and ERA5 (0.25$^\circ$) for ocean \\
Pmean ($-$) & Mean annual $P$ & CHELSA V2.1 (1-km resolution; \citealp{karger_climatologies_2017}) \\
ETH (m) & Effective Terrain Height (ETH) calculated following \citet{daly_physiographically_2008} & Global Multi-resolution Terrain Elevation Data (GMTED) 2010 \citep{danielson_global_2011} \\
Lat ($^\circ$) & Latitude & / \\
Lon ($^\circ$) & Longitude & / \\
AbsLat ($^\circ$) & Absolute latitude & / \\
\bottomrule
\end{tabularx}

\label{StatPredictor}
\end{table}

\section*{Daily and Hourly Gauge Observations}
\label{DataPrecipGaugeObs}

We compiled a vast worldwide archive of hourly and daily $P$ gauge observations for; (i)~training and validating the ML models used to generate the baseline $P$ fields (Appendix~A ``\nameref{MethodsMLalgorithm}''); and (ii)~correcting those baseline fields (using daily data only; Appendix~A ``\nameref{MethodsDailyGaugeCorrection}'').

For the ML model training and validation (Appendix~A ``\nameref{MethodsMLalgorithm}''), we used both direct $P$ gauge observations and gridded gauge-radar $P$ data. Direct gauge observations were obtained from the Global Historical Climatology Network-Daily (GHCN-D; \citealp{menne_overview_2012}; $n=127,558$) and the Global Summary of the Day (GSOD; \citealp{noaa_ncep_global_2020}; $n=25,640$), supplemented with national databases for Brazil ($n=12,410$), Mexico ($n=5,398$), Peru ($n=255$), Iran ($n=3,122$), and Saudi Arabia ($n=460$). Hourly data were taken from the Global Sub-Daily Rainfall archive (GSDR; \citealp{lewis_gsdr_2019}; $n=20,633$) and the Integrated Surface Database (ISD; \citealp{noaa_ncei_global_2021}; $n=25,937$).

For Europe and the contiguous U.S., we used gridded gauge-radar datasets: EUropean RADar CLIMatology (EURADCLIM; hourly, 2~km; 2013--2022; \citealp{overeem_euradclim_2023}) and NCEP Stage-IV (hourly, 4~km; 2002--present; \citealp{lin_12_2005}), respectively. These datasets were aggregated to daily totals and upscaled to 0.1$^\circ$ using averaging to match the spatial resolution of MSWEP~V3. To maximize the quality of the training data, we only used the gauge-radar time series at gauge locations ($n=16,497$ for EURADCLIM and $n=20,635$ for Stage-IV).

Where available, gridded gauge-radar datasets are preferred over point-based gauge observations for ML model training and validation for three reasons. Firstly, they provide grid-cell averages with probability distributions (i.e., $P$ frequencies and peak magnitudes) matching those of MSWEP~V3, thereby reducing the point-to-grid scale mismatch \citep{yates_point_2006,ensor_statistical_2008}. Secondly, they are reported in UTC, eliminating spurious temporal shifts between daily totals from the observations and the baseline fields (\citealp{yang_development_2020}; see also Appendix~A ``\nameref{MethodsReportingTimes}''). Thirdly, and finally, these products typically undergo more rigorous quality control procedures to exclude erroneous data.

For the daily correction of the baseline $P$ fields (Appendix~A ``\nameref{MethodsDailyGaugeCorrection}''), only direct gauge observations were used, since the gauge-radar datasets have limited temporal coverage (EURADCLIM: 2013--2022; Stage-IV: 2002--present).

\section*{Daily Gauge Data Quality Control}
\label{DailyGaugeDataQC}

Time-series of each daily $P$ were screened to remove suspicious gauges. We identified and removed erroneous zero and non-zero $P$ sequences \citep{durre_comprehensive_2010,funk_climate_2015}. This procedure used a central annual moving average and a moving minimum, requiring at least six months of data. Values were retained only when the moving average was non-zero and the moving minimum was zero.

The following criteria were subsequently applied to determine gauge inclusion:
\begin{enumerate}
\item Daily maximum $P$ must not exceed $1{,}825$~mm~d$^{-1}$, the highest daily rainfall ever recorded (\url{www.weather.gov/owp/hdsc_world_record}), to discard records with physically implausible extremes.
\item Long-term mean $P$ must lie between 5 and 10,000~mm~yr$^{-1}$, to exclude records with unrealistic climatologies due to unit/aggregation or data corruption errors.
\item The record must include at least five days with $P > 1$~mm~d$^{-1}$, to ensure sufficient non-trivial wet events for robust fitting and evaluation.
\item At least 20\% of days must have $P < 0.5$~mm~d$^{-1}$, to preserve a realistic fraction of dry/light-$P$ days.
\item At least 10\% of days must have $P < 0.001$~mm~d$^{-1}$, again to preserve a realistic fraction of dry/light-$P$ days.
\item The record must contain at least one full year of valid daily data for ML model training/validation or at least five years for the baseline correction.
\item For gauges within 2-km radius, only the highest-priority source was retained (EURADCLIM $=$ Stage-IV $>$ GHCN-D $>$ GSDR $>$ ISD $>$ national databases $>$ GSOD).
\end{enumerate}

For ML model training and validation (2010--2024; Appendix~A~``\nameref{MethodsMLalgorithm}'' and subsection~\ref{DataMethods}\ref{methodsbaselinevalidation}), a total of 31,917 gauges (direct and radar-gauge) passed all quality control criteria. Of these, 50\% (15,959) were randomly assigned for model training and 50\% (15,958) for validation purposes. For the daily gauge-correction procedure (1979--2024; Appendix~A~``\nameref{MethodsDailyGaugeCorrection}''), 57,666 gauges (only direct) were retained.

Hourly gauge observations (from GSDR, Stage-IV, and EURADCLIM) were not subjected to additional quality control, as these datasets already undergo rigorous checks prior to release.

\section*{Monthly Gauge Observations}
\label{MonthlyPrecipData}

In addition to the daily correction of the baseline fields using daily gauge observations (Appendix~A ``\nameref{MethodsDailyGaugeCorrection}''), we also carried out a monthly correction of the baseline, leveraging the better availability of monthly gauge data. For this purpose, we used the comprehensive gridded GPCC Full Data Monthly Product Version 2022 \citep{schneider_evaluating_2017}, which incorporates approximately 86,000 gauges and spans 1891--2020 at 0.25$^\sim$ resolution. We used the GPCC dataset rather than direct monthly gauge observations because GPCC provides the most comprehensive global compilation of monthly gauge data currently available. To extend coverage to the present, we used the GPCC First Guess Monthly Product (2013--present), available at 0.5$^\circ$ resolution with a latency of approximately five days. Because the two GPCC products differ in spatial resolution (0.25$^\circ$ vs.\ 0.5$^\circ$) and baseline climatologies (GPCC Climatology V2015 vs.\ V2022) they cannot be directly concatenated. We addressed this by: (i)~downscaling the First Guess data to 0.25$^\circ$ via nearest-neighbor interpolation; and (ii)~rescaling the First Guess values to match the Full Data mean at each 0.25$^\circ$ grid cell for the period of overlap (2013--2020).

To reduce uncertainty, we retained only those GPCC grid cells that contain actual gauge observations, discarding cells representing purely interpolated values. Thus, we essentially reverse-engineered the original monthly observations from the gridded data. Where multiple gauges were present in a single grid-cell, we treated them as a single gauge observation. A limitation of this approach is that the exact gauge locations are unknown; only their presence within each 0.25$^\circ$ grid cells is indicated. No further quality control was applied to the GPCC data, as the underlying gauge observations have already undergone rigorous quality control by the developers \citep{schneider_evaluating_2017}.

\section*{Baseline Precipitation Estimation Algorithm}
\label{MethodsMLalgorithm}

MSWEP~V3 applies the algorithm of \citet{wang_saudi_2025} to produce global baseline $P$ fields. The algorithm employs stacked decision tree-based models that integrate dynamic predictors related to $P$ and $T$ (from ERA5, IMERG, GSMaP, and PERSIANN-CCS-CDR) and static predictors describing geographic location, climate, and topography (Appendix~A ``\nameref{DataPredictors}''). Each stack consists of four submodels, individually optimized for different temporal resolutions or for correcting variance underestimation (see Fig.~\ref{Flowchart}).

The first submodel is an XGB model \citep{chen_xgboost_2016} designed to estimate daily $P$. It is trained on global daily $P$ observations pooled together (Appendix~A ``\nameref{DataPrecipGaugeObs}''), leveraging their broader spatial and temporal availability compared to hourly observations. The second submodel, also an XGB model, estimates 3-hourly $P$ and is used to temporally disaggregate the daily estimates. It is trained using hourly observations aggregated to 3-hourly resolution (Appendix~A ``\nameref{DataPrecipGaugeObs}''). The third submodel, an RF model \citep{breiman_random_2001}, corrects the variance underestimation typical of ML-based $P$ estimates (e.g., \citealp{he_spatial_2016}). It uses the disaggregated 3-hourly $P$ estimates from the second submodel together with the static predictors, and is trained using 3-hourly observations. For training, both the disaggregated estimates and the observations are independently sorted for each gauge over their period of overlap to construct quantile-matching pairs. These sorted pairs are then pooled across all gauges to train the model. To improve the correction of wet-day frequencies and low-intensity events, the disaggregated estimates are square-root transformed prior to training and inference, and squared afterwards. The fourth submodel, another XGB model, estimates hourly $P$ by further disaggregating the variance corrected 3-hourly outputs from the third submodel. All submodels except the third incorporate both dynamic and static predictors. The third submodel uses only static predictors along with the disaggregated $P$ estimates.

The MSWEP~V3 baseline is produced using multiple model stacks to account for the variable availability of dynamic predictors across space and time. These variations arise, for example, from the absence of satellite data at high latitudes ($>60^\circ$N), the unavailability of IMERG and GSMaP before 2000, and the five-day delay in ERA5 availability (Fig.~\ref{Periods}a). In total, 18 model stacks were trained, each incorporating a different combination of dynamic predictors (Table~\ref{model_predictors_table}) together with the static predictors. The stacks are ordered broadly from most to least accurate, with lower numbers generally corresponding to combinations that include a larger set of higher-quality dynamic predictors. model\_01, which incorporates ERA5, IMERG, and GSMaP, is listed first because it serves as the reference stack; model\_05, which uses ERA5 only, serves as the reference at high latitudes where satellite data are unavailable or unreliable. For each 0.1$^\circ$ grid cell and hourly time step, the production pipeline identifies all applicable model stacks based on predictor availability and applies the lowest-numbered eligible stack.

Because different model stacks produce $P$ estimates with distinct error and bias characteristics, combining their predictions can introduce artificial temporal discontinuities. To avoid this, the historical MSWEP~V3 production pipeline first harmonizes the outputs of all non-reference stacks to the reference stack (model\_01 or model\_05 at high latitudes) before combining them (Fig.~\ref{Periods}b). The harmonization procedure consists of three steps: (i)~detrending each time series by dividing it by a moving annual mean; (ii)~applying a CDF matching to model\_01 (if available) or otherwise to model\_05, which relies only ERA5 and is always available for the historical production; and (iii)~re-scaling the adjusted series using the same moving annual mean to restore the original trend. The detrending step ensures that the CDF matching adjusts only the distributional shape without distorting underlying long-term trends.

We used gauge observations and predictor data between 2010--2024 for training and validation. Details on gauge quality control and the training-validation split are provided in Appendix~A ``\nameref{DailyGaugeDataQC}''. To avoid temporal inconsistencies when training the first submodel, daily totals (for $P$) and daily means (for $T$) of the dynamic predictors were computed over the 24-hour period ending at each gauge's inferred reporting time (Appendix~A ``\nameref{MethodsReportingTimes}''). For the second and third submodels, hourly gauge time series were shifted according to each station's inferred reporting-time offset so that all timestamps align with UTC (Appendix~A ``\nameref{MethodsReportingTimes}''). The sources of daily and hourly gauge observations are described in Appendix~A ``\nameref{DataPrecipGaugeObs}''. RF models were implemented in Python with \texttt{scikit-learn}; XGB models with \texttt{xgboost}. Hyperparameters are listed in Table~\ref{table:xgboost_hyperparameters} (XGB) and Table~\ref{table:rf_hyperparameters} (RF).

\begin{table}[h!]
\centering
\caption{Hyperparameters for the XGBoost models.}
\label{table:xgboost_hyperparameters}
\begin{tabular}{lc}
\hline
Hyperparameter    & Value \\ \hline
\texttt{n\_estimators}     & 100            \\ \texttt{max\_depth}        & 12             \\ \texttt{min\_child\_weight} & 5             \\ \texttt{colsample\_bytree} & 0.7            \\ \texttt{gamma}             & 2              \\ \texttt{reg\_alpha}        & 0.5            \\ 
\texttt{reg\_lambda}       & 0.5            \\ 
\texttt{learning\_rate}    & 0.2            \\ \hline
\end{tabular}
\end{table}

\begin{table}[h!]
\centering
\caption{Hyperparameters for the RF models.}
\label{table:rf_hyperparameters}
\begin{tabular}{lc}
\hline
Hyperparameter    & Value \\ \hline
\texttt{n\_estimators}     & 100            \\ 
\texttt{max\_depth}        & 15             \\ 
\texttt{min\_samples\_split} & 5            \\ 
\texttt{min\_samples\_leaf} & 5            \\ 
\texttt{max\_features}     & 0.7            \\ \hline
\end{tabular}
\end{table}

\section*{Near Real-Time Baseline Extension}
\label{MethodsNRTextension}

MSWEP-NRT extends the historical MSWEP~V3 record to the present with a latency of approximately two hours, incorporating data from GDAS \citep{noaa_ncep_national_2025}, IMERG \citep{huffman_gpm_2019}, GSMaP \citep{kubota_global_2020}, and PDIR-Now (\citealp{nguyen_persiann_2020}; Table~\ref{globalPProductDetails}). While the historical MSWEP~V3 production pipeline will be run infrequently (likely a few times per year) and is designed to ensure time-series consistency at the grid-cell level (Appendix~A ``\nameref{MethodsMLalgorithm}''), the MSWEP-NRT production pipeline is executed automatically multiple times per hour as new dynamic input data become available, generating $P$ estimates for each hourly timestep with new data, and becoming fully consistent with the historical record once ERA5 reanalysis data are released.

Similar to the baseline model, the ML model stack applied for each time step and grid cell also depends on the availability of dynamic predictors (Table~\ref{model_predictors_table}). As new data sources arrive, the system applies a model stack with a lower index number and higher-quality inputs, progressively refining the NRT estimates. The earliest estimates rely solely on model\_18, which uses the purely infrared-based PDIR-Now. Although PDIR-Now has the shortest latency (approximately two hours), it has low accuracy compared to the other products. As GDAS analysis data, and IMERG and GSMaP microwave $P$ estimates become available, models\_15, 10, 11, and 07 are used to produce updated, more accurate estimates. Once ERA5 becomes available (typically after about five days), model\_01 is applied at low and mid-latitudes and model\_05 at high latitudes---the same model stacks used as reference to derive the historical baseline (Appendix~A ``\nameref{MethodsMLalgorithm}''). At this stage, the NRT estimates are finalized and become fully consistent with the historical estimates. Each hourly MSWEP-NRT file includes a indicator field identifying the model stack applied (1--18), ensuring full traceability. 

The first, second, and third submodels in each ML stack operate at daily, 3-hourly, and 3-hourly resolutions, respectively. Therefore, they require complete 24- or 3-hour input sequences, which are not yet available for the current day and hour. To address this, the NRT production pipeline identifies the most recent hourly file for each data source, constructs 24-hour input windows ending at the most recent complete timestamp per source, and generates interim estimates for the ongoing day. These estimates are provisional and replaced as new data become available and the full day sequence is completed.

\section*{Estimation of Gauge Reporting Times}
\label{MethodsReportingTimes}

In most countries, daily gauge $P$ amounts are measured and reported each morning in local time, traditionally after the observer wakes. For example, in Italy, Poland, and the UK this generally occurs around 08:00 or 09:00 local time (06:00--08:00 UTC; \citealp{becherini_adjustment_2024,twardosz_influence_2011}). These reporting times are variously referred to as the daily reporting time \citep{beck_mswep_2019,xiang_evaluation_2021}, time stamp \citep{contractor_rainfall_2020}, boundary time \citep{yang_development_2020,wang_saudi_2025}, measurement interval \citep{overeem_euradclim_2023}, end of day (EOD; \citealp{yatagai_end_2020}), or $P$ day \citep{twardosz_influence_2011}. Reporting times vary substantially across time zones, countries, and regions, depending on local observational practices \citep{yang_development_2020}. Even automated gauges, such as those from the GSOD database, though officially reporting totals for 00:00--24:00~UTC, in practice have effective reporting times several hours earlier \citep{beck_mswep_2019}, as the measurements still include $P$ from the previous day \citep{noaa_ncep_global_2020}. 

Accounting for reporting times is crucial to avoid temporal misalignment when comparing daily totals from satellite and (re)analysis products with those from gauges. In the UK, for instance, ignoring the local reporting time (08:00~UTC) would result in a 33\% chance ($100\times 8/24$) that a one-hour $P$ event is assigned to the wrong UTC day, degrading both ML model training (Appendix~A ``\nameref{MethodsMLalgorithm}'') and gauge correction (Appendix~A ``\nameref{MethodsDailyGaugeCorrection}''). Therefore, similar to \citet{beck_mswep_2019} and \citet{overeem_euradclim_2023}, reporting times were inferred for each daily gauge using the following procedure: (i)~the hourly baseline $P$ data were shifted in 1-hour increments from $-36$ to $+36$ hours; (ii)~daily $P$ totals for 00:00 to 24:00 UTC were recalculated for each shift; (iii)~daily Spearman correlations (instead of Pearson correlations, because Spearman is rank-based and less dominated by large events) were computed between each shifted version and the gauge data; (iv)~the shift yielding the highest correlation was identified, and the negative of this optimal shift was taken as the reporting time for the gauge.

Similarly, hourly gauge observations often reflect local time or include unknown timestamp offsets, potentially impairing the training of the 3-hourly and hourly submodels (Appendix~A ``\nameref{MethodsMLalgorithm}''). To address this, each gauge time series was shifted in 1-hour increments from $-36$ to $+36$ hours, and the shift that maximized correlation with the MSWEP~V3 baseline was applied to align the timestamps with UTC.

\section*{Estimation of Precipitation Correlation Lengths}
\label{MethodsCorrelationLengths}

The correlation length of $P$ quantifies the spatial scale over which $P$ values remain statistically dependent \citep{mandapaka_analysis_2013}. It varies with $P$ type and temporal resolution: stratiform $P$ and longer accumulation periods typically yield longer correlation lengths, while convective $P$ and shorter durations result in shorter ones. Herein, we define correlation length as the e-folding distance of the spatial correlation function, i.e., the distance at which the correlation decays to $1/e$ (approximately 0.37) of its value at zero separation. For example, gauge-based analyses reported daily correlation lengths of 102--202~km across mainland China \citep{fan_spatial_2021} and around 94~km in Belgium \citep{ly_geostatistical_2011}, both regions dominated by stratiform $P$. In contrast, substantially shorter hourly and daily distances---about 10~km and 33~km, respectively---were observed in Singapore \citep{mandapaka_analysis_2013}, where convective $P$ prevails. 

Consideration of $P$ correlation lengths is critical when using gauge data to correct radar, satellite, or (re)analysis $P$ fields, as these lengths define the effective spatial influence of each gauge. Several studies have implemented such corrections using empirically derived correlation lengths, including applications in Switzerland \citep{schiemann_geostatistical_2011}, the Netherlands \citep{schuurmans_automatic_2007}, and Mexico \citep{tapia-silva_geostatistics_2024}. Notably, the CHIRPS~V2 quasi-global $P$ dataset incorporates correlation lengths into its five-day gauge correction procedure \citep{funk_climate_2015}.

In MSWEP~V3, daily correlation lengths were used as inputs to the daily and monthly OI baseline gauge correction schemes (Appendix~A ``\nameref{MethodsDailyGaugeCorrection}'' and ``\nameref{MethodsMonthlyGaugeCorrection}'', respectively). A daily correlation length was estimated for each 0.1$^\circ$ grid cell from the baseline $P$ fields by: (i)~randomly selecting 20 neighboring grid cells within a 500-km radius (only 20 neighbors to reduce computational demand); (ii)~computing 20 pairwise Spearman correlations (instead of Pearson correlations, because Spearman is rank-based and less dominated by large events) between the daily baseline $P$ series (2000--present) of the target grid cell and each selected neighbor; and (iii)~fitting a Gaussian decay model (Eq.~\ref{EqGaussianCorrelationModel}). Only baseline $P$ data from 2000 onward were used, as data quality is higher during this period.

In practice, correlation lengths vary not only spatially but also temporally, exhibiting seasonal shifts driven by changes in $P$ type, intensity, and atmospheric dynamics \citep{mandapaka_analysis_2013,zhang_impact_2017,kursinski_spatiotemporal_2008}. For example, \citet{tokay_experimental_2014} found strong seasonal contrasts in Delaware, with 30-minute correlation lengths of 8–13~km in summer and 51–85~km in winter. However, to maintain temporal consistency in the gauge-corrected MSWEP~V3 series, we adopted a fixed correlation length. Allowing it to vary over time would introduce temporal variability in gauge influence, compromising homogeneity.

\section*{Daily Gauge Correction Procedure}
\label{MethodsDailyGaugeCorrection}

A wide range of methods have been developed to correct satellite or reanalysis $P$ fields using rain gauge observations. These approaches span from simple additive or multiplicative bias corrections (e.g., \citealp{habib_effect_2014}) to more advanced techniques, including Barnes' objective analysis (e.g., \citealp{overeem_euradclim_2023}), inverse distance weighting (IDW; e.g., \citealp{boushaki_bias_2009}), optimal interpolation (OI; e.g., \citealp{pan_analysis_2012}), kriging (e.g., \citealp{verworn_spatial_2011}), and ML (e.g., \citealp{baez-villanueva_rf-mep_2020}). In MSWEP~V3 we apply OI \citep{gandin_objective_1963,daley_atmospheric_1993,chen_global_2002} to correct the daily variability of hourly baseline $P$ fields (see Appendix~A ``\nameref{MethodsMLalgorithm}'') using daily gauge observations (Appendix~A ``\nameref{DataPrecipGaugeObs}''). The method is computationally efficient---crucial given the dataset's global hourly coverage at 0.1$^\circ$ resolution from 1979 to the present---and accounts for spatial gauge proximity, $P$ correlation lengths, and inter-gauge dependencies, providing a statistically grounded weighting of each gauge's influence on its surroundings. Our implementation also accounts for reporting times to avoid temporal mismatches in the 24-hour accumulation period between the baseline and gauge observations (see Appendix~A ``\nameref{MethodsReportingTimes}''). The correction process is applied independently to each grid cell (the ``target'') as:
 
\begin{enumerate}
\item Identify the nearest gauge within a 500-km radius in each of four quadrants (0–90°, 90–180°, 180–270°, and 270–360°) around the center of the target, measured clockwise from true north, to maximize independence among selected gauges.

\item Add small random values to the hourly baseline time series to prevent zero-value conflicts when the gauge reports $P$ but the baseline does not.

\item Rescale both the gauge observations and the baseline time series at the surrounding gauges to match the target baseline grid cell over their overlapping periods. This reduces the likelihood of introducing shifts in $P$ characteristics when gauge records start or stop.

\item Generate corrected hourly target baseline series using gauge and baseline data from each surrounding gauge as follows:

\begin{enumerate}
\item Shift each surrounding hourly baseline series by the inverse of its estimated reporting time to ensure temporal alignment with the daily gauge data.

\item Compute daily correction amounts for each surrounding gauge by differencing the daily totals between the gauge and its shifted baseline, then apply these corrections to the target's shifted daily baseline.

\item Compute daily correction factors by dividing the corrected daily target baseline by the uncorrected daily target baseline.

\item Disaggregate the daily correction factors and apply them multiplicatively to the shifted hourly target baseline. Finally, apply the reverse shift to restore the original temporal alignment.

\end{enumerate}

\item Compute vector of correlations based on distances between the target location and each surrounding gauge ($\mathbf{C}_{to}$), and matrix of correlations based on the inter-gauge distances ($\mathbf{C}_{oo}$), using a Gaussian spatial correlation model:
\begin{equation}
\rho(d) = \exp\left(-\frac{d^2}{L^2}\right),
\label{EqGaussianCorrelationModel}
\end{equation}
where $d$ (km) is the distance between locations and $L$ (km) is the spatial correlation length (i.e., e-folding distance, see Appendix~A ``\nameref{MethodsCorrelationLengths}'').

\item Use OI to compute the weight vector $\mathbf{w}$ as follows:
\begin{equation}
\mathbf{w} = \mathbf{C}_{to} \left(\mathbf{C}_{oo} + \gamma \mathbf{I}\right)^{-1},
\end{equation}
where $\gamma$ is the observation-to-background error variance ratio, and $\mathbf{I}$ is the identity matrix. We tested $\gamma \in {0.05, 0.1, 0.3}$ and found that $\gamma=0.05$ yielded the best performance, although the differences were small. A pseudoinverse is used for matrix inversion to maintain numerical stability in the presence of ill-conditioned and/or incomplete correlation matrices.

\item Merge the corrected hourly target baseline time series using the OI-based weights, with the residual weight assigned to the uncorrected hourly target baseline, as follows:
\begin{equation}
\hat{P}_{t} = \frac{\sum_{i=1}^{N+1} w_i \cdot P_{i,t}}{\sum_{i=1}^{N+1} w_i \cdot \mathbb{I}_{i,t}},
\end{equation}
where $P_{i,t}$ is the $i$th corrected hourly estimate (or the uncorrected baseline for $i=N+1$), $w_i$ is the corresponding weight, and $\mathbb{I}_{i,t}$ is an indicator for data availability (1~if valid, 0~if missing). 
    
\end{enumerate}

\section*{Monthly Gauge Correction Procedure}
\label{MethodsMonthlyGaugeCorrection}

Many global gridded $P$ products apply monthly corrections using gauge-based products such as GPCC \citep{schneider_evaluating_2017}. The typical approach---used, for example, in IMERG-Final \citep{huffman_gpm_2019}, WFDE5 \citep{cucchi_wfde5_2020}, GPCP \citep{huffman_new_2023}, and PERSIANN-CCS-CDR \citep{sadeghi_persiann-ccs-cdr_2021}---involves adjusting the monthly totals of the baseline $P$ field to match GPCC's gridded values directly. However, this approach treats all grid cells equally, including those far from any actual gauge, where interpolated values are subject to considerable uncertainty. It also ignores the spatial correlation length of $P$, leading to suboptimal adjustments.

While we still use GPCC data to correct MSWEP~V3, we applied a more sophisticated approach. Firstly, we restrict the use of GPCC data to grid cells with direct gauge observations, excluding purely interpolated values (as detailed in Appendix~A ``\nameref{MonthlyPrecipData}''). Secondly, we apply an OI scheme nearly identical to that used for the daily correction (see Appendix~A ``\nameref{MethodsDailyGaugeCorrection}''), to account for spatial $P$ correlation structure, inter-gauge dependencies, and gauge proximity, allowing each gauge to influence its surroundings in a statistically defensible way. The scheme was applied on a 0.1$^\circ$ grid-cell basis after completion of the daily gauge correction. We used correlation lengths estimated from daily $P$ data (see Appendix~A ``\nameref{MethodsCorrelationLengths}''), as monthly correlations are inflated by temporal smoothing and do not reflect the scale of individual $P$ events. Daily reporting times were disregarded as they have negligible impact at the monthly scale. 

Additionally, we applied a long-term bias adjustment, as our OI implementation only corrects temporal variability. We used monthly rather than daily observations for this adjustment due to the greater availability of monthly gauge data worldwide \citep{schneider_new_2016}. For each 0.1$^\circ$ grid-cell, the adjustment involved: (i)~computing the absolute long-term bias (mm month$^{-1}$) at the four nearest gauges in each quadrant by subtracting the gauge mean from the baseline mean for the period of overlap (provided it is $>5$~years long); (ii)~estimating the long-term bias at the target location using a weighted average of these biases, with weights from the monthly OI application; (iii)~subtracting the estimated bias from the long-term mean of the corrected baseline at the target location; and (iv)~rescaling the hourly corrected baseline series to match this adjusted long-term mean. 

\section*{Validation Metrics}
\label{validationmetrics}

The Kling-Gupta Efficiency (KGE) is defined as:
\begin{equation}
\textrm{KGE} = 1 - \sqrt{(r - 1)^2 + (\beta - 1)^2 + (\gamma - 1)^2},
\end{equation}
where $r$ is the Pearson correlation coefficient, $\beta$ is the bias ratio (mean estimated divided by mean observed), and $\gamma$ is the variability ratio (estimated standard deviation divided by observed).

The Critical Success Index (CSI), evaluated for events exceeding 10~mm~d$^{-1}$, is given by:
\begin{equation}
\textrm{CSI} = \frac{H}{H + M + F + 10^{-9}},
\end{equation}
where $H$ denotes hits (correctly estimated events), $M$ denotes misses (unpredicted events), and $F$ denotes false alarms (incorrectly predicted events). A small constant prevents division by zero.

Peak bias at the 99.5th percentile ($B_\textrm{peak}$; \%) is calculated as:
\begin{equation}
B_\textrm{peak} = 100 \times \frac{P_{99.5} - O_{99.5}}{O_{99.5}},
\end{equation}
where $P_{99.5}$ and $O_{99.5}$ are the 99.5th percentiles of the estimated and observed values, respectively.

Wet day bias ($B_\textrm{wet,days}$; days) quantifies the difference in the number of wet days (defined as days exceeding 0.5~mm~d$^{-1}$) as:
\begin{equation}
B_\textrm{wet,days} = 365.25 \times \frac{P - O}{N},
\end{equation}
where $P$ and $O$ are the number of wet days in the estimated and observed time series, and $N$ is the total number of time steps.

\acknowledgments
We would like to express our gratitude to the creators of the $P$ products, as this research would not have been possible without them. Further, we highly appreciate the authors of the Python modules that proved instrumental to this study, specifically \texttt{numpy} \citep{van_der_walt_numpy_2011}, \texttt{scipy} \citep{virtanen_scipy_2020}, \texttt{scikit-image} \citep{walt_scikit-image_2014}, \texttt{matplotlib} \citep{hunter_matplotlib_2007}, \texttt{pandas} \citep{mckinney_data_2010}, and \texttt{scikit-learn} \citep{pedregosa_scikit-learn_2011}. 

%
%
\datastatement

MSWEP~V3 is available for download at \url{www.gloh2o.org/mswep}. CPC Unified is available on the
NOAA Physical Sciences Laboratory (PSL) website (\url{https://psl.noaa.gov/data/gridded/data.cpc.globalprecip.html}). IMERG can be accessed from the NASA Global Precipitation Measurement (GPM) website (\url{https://gpm.nasa.gov/data}). JRA-3Q is available via the National Center for Atmospheric Research (NCAR) Research Data Archive (RDA; \url{https://rda.ucar.edu/datasets/ds640000/dataaccess}). GPCP is accessible via the NOAA PSL website (\url{https://psl.noaa.gov/data/gridded/data.gpcp.html}). SM2RAIN-ASCAT, SM2RAIN-CCI, and GPM+SM2RAIN are 
hosted on Zenodo (\url{https://zenodo.org/records/10376109}, \url{https://zenodo.org/records/1305021}, and \url{https://zenodo.org/records/3854817}, respectively). ERA5 data can be obtained from the Copernicus Climate Data 
Store (CDS; \url{https://cds.climate.copernicus.eu/datasets/reanalysis-era5-single-levels?tab=overview}). CHIRP and 
CHIRPS are available via the University of California Climate Hazards Center (CHC) website 
(\url{www.chc.ucsb.edu/data/chirps/}). PERSIANN-CCS-CDR and PDIR-Now are accessible via the Center for Hydrometeorology and Remote Sensing (CHRS) website (\url{https://chrsdata.eng.uci.edu/}). All websites last accessed 15 December, 2025.


%





%



\bibliographystyle{ametsocV6}
\bibliography{references}

\end{document}